# Gauge symmetry breaking, collective modes and boson superconductivity in the $t$-$J$ model[*]

Maxim Raykin

*Physics Department, Boston University, Boston, MA 02215*

August 19, 1994

**Abstract**

A theory of the $t$-$J$ model in the presence of an external electromagnetic field is presented. The $1/N$ expansion for this model in the slave-fermion representation is developed, and it is shown that all the general properties of the theory are satisfied in every order of the $1/N$ expansion separately. A convenient procedure of the gauge fixing is suggested. It is shown that superconductivity may exist in this model even in the absence of the electron pairing. It is argued that experimental predictions of the theory might agree with the observed properties of the copper oxides.

## I. Introduction

Soon after the discovery of superconductivity in the rare earth-based copper oxides [1], Anderson [2] suggested that the physics of these materials may be described by a large-$U$ Hubbard model and the related $t$-$J$ model. The Hamiltonian of the $t$-$J$ model operates in a constrained subspace which excludes double occupancy. To cope with the constraint, the electron operators are usually represented as

$$\psi^\dagger_{\sigma\mathbf{r}} = b^\dagger_{\sigma\mathbf{r}} a_{\mathbf{r}}. \tag{1.1}$$

The operator $a_\mathbf{r}$ here represents an empty site, while the operator $b_{\sigma\mathbf{r}}$ represents the electron's spin as $\mathbf{S}_\mathbf{r} = \frac{1}{2} b^\dagger_{\alpha\mathbf{r}} \vec{\sigma}_{\alpha\beta} b_{\beta\mathbf{r}}$. In terms of these operators, the constrained subspace is selected by the holonomic condition

$$b^\dagger_{\sigma\mathbf{r}} b_{\sigma\mathbf{r}} + a^\dagger_\mathbf{r} a_\mathbf{r} = 1, \tag{1.2}$$

which has to be imposed independently at every site $\mathbf{r}$. Representation (1.1) simply expresses the fact that the states of the electron system with forbidden double occupancy may be put into one-to-one correspondence with the states of the $b$-$a$ system, satisfying condition (1.2). The corresponding map is given by Eq. (1.1). To preserve the fermionic statistics of electrons, we must assign bosonic statistics to one of the particles $a$ or $b$ and fermionic statistics to the other. Either possibility is admissible, and we choose to assign the fermionic statistics to $a$; the particles $b$ then are usually called Schwinger bosons (SB).

Representation (1.1) corresponds to the separation of spin and density degrees of freedom [3]. Their excitations – spinons and holons – rather then electrons, have been claimed to be the real quasiparticles in

---

the copper oxides. If this is true, then what might be the nature of the superconductivity in the system of interacting spinons and holons?

The phenomenon of superconductivity may be described as an emergence of the photon's mass due to the Anderson-Higgs mechanism. The Higgs field, which screens the photon, is necessarily bosonic; to create such a field, electrons in the conventional BCS theory have to form a bound state – a Cooper pair. On the other hand, in the spin-density separation approach to the copper oxides, we have the boson field $b_\sigma$ as one of the *fundamental* fields of the theory. Then the natural question to ask is whether this field can screen an external electromagnetic field (EMF), leading to superconductivity *without* electron pairing?

It is clear that the only way to answer this question is to develop a reliable approach to the $t$-$J$ model in the presence of an EMF, and to check whether an external *magnetic* field in this model can create a nonvanishing electric current in the phase having unpaired electrons. This program is realized in the present paper, and we demonstrate that the question raised above has a positive answer. We note that the spin-density separation cannot be considered a trivial consequence of representation (1.1). Indeed, this representation naturally introduces to the theory invariance with respect to the local gauge transformation $a_{\mathbf{r}}(b_{\sigma\mathbf{r}}) \to e^{i\alpha_{\mathbf{r}}} a_{\mathbf{r}}(b_{\sigma\mathbf{r}})$. Thus the theory becomes a lattice gauge theory [4], where the presence of the fields $b_\sigma$ and $a$ in the initial Lagrangian does not automatically guarantee their existence as real particles. Indeed, if the gauge field is in the confining phase, it binds spinons and holons together to form a physical electron as a bound state. The electron then carries both spin and density degrees of freedom, so that in effect there is no spin-density separation.

The real spin-density separation occurs as a result of spontaneous gauge symmetry breaking with the transition to the phase with nonzero average gauge field and unconfined slaves $b_\sigma$ and $a$ (deconfinement transition). That kind of transition was discovered on the mean-field (MF) level in the very first works on lattice gauge theory [5, 6]. Soon, however, Elitzur [7] proved that in the lattice gauge theory the average value of the gauge field must vanish. The reconciliation, achieved in numerous later works, may be described as follows. Initially, one does not need to fix the gauge in lattice gauge theory [5], because the gauge group there is compact. In such a theory without gauge fixing and with manifest gauge invariance, the expectation values of the gauge-noninvariant quantities do indeed vanish due to the Elitzur theorem. On the other hand, one can fix the gauge, using the Faddeev-Popov trick, and factor out the trivial volume of the gauge group. This procedure (which is also necessary for the development of the diagram technique) will not affect the physical, gauge invariant results. Fixing the gauge, however, spoils the gauge invariance of the theory. The Elitzur theorem is not applicable to it anymore, and the gauge-noninvariant quantities can obtain now nonvanishing expectation values. Therefore, on one hand, the lattice gauge theory with the fixed gauge gives correct values of the gauge-invariant quantities, and on the other hand, allows one to give a convenient description of the different phases of the theory. Indeed, here we can use for that description the expectation values of the gauge-noninvariant quantities, such as the gauge field itself. This is more convenient than the use of gauge-invariant (and more complicated) objects, such as the Wilson loop [4, 5], in the theories without gauge fixing.

In this paper we will follow this route in the study of spontaneous gauge symmetry breaking and spin-density separation. The outline of the paper is as following: we start in the next section with the derivation of the diagram technique for the $t$-$J$ model. We develop a systematic $1/N$ expansion, and demonstrate throughout the paper that all general properties of the theory are satisfied in each order of this expansion separately. The zeroth order, corresponding to the $N \to \infty$ limit, yields the mean-field theory (MFT). We work consistently on the lattice without the transition to the continuous limit and suggest a convenient procedure of the gauge fixing. In the following sections we discuss the general properties of the theory: Sec. III contains the discussion of collective modes, which exist in the phases with a broken gauge symmetry, while Sec. IV shows the relation between the diagram technique and the equations of motion, following from the Lagrangian. The interaction with an EMF is introduced in Sec. V. In Sec. VI we demonstrate that this interaction satisfies the Ward identities. The superconducting phase is considered in Sec. VII, and finally, in Sec. VIII, we discuss the possibility of describing the experimentally observed properties of the copper oxides in the framework of our approach. Since this paper is devoted mainly to the formulation and the basic consideration of our approach to the $t$-$J$ model, the explicit comparison between our theory and experiment will not be reported here.

## II.  $1/N$ expansion and diagram technique

### II.1  Basic definitions

Our starting point is the $t$-$J$ model, which can be represented by the Hamiltonian [8]

$$H = \sum_{\mathbf{r}\eta} \left[ -t_\eta \left( \psi^\dagger_{\sigma\mathbf{r}} \psi_{\sigma\mathbf{r}+\hat{\eta}} + \psi^\dagger_{\sigma\mathbf{r}+\hat{\eta}} \psi_{\sigma\mathbf{r}} \right) - \frac{J_\eta}{2} C^\dagger_{\eta\mathbf{r}} C_{\eta\mathbf{r}} \right], \tag{2.1}$$

where $\psi^\dagger_{\sigma\mathbf{r}}$ ($\psi_{\sigma\mathbf{r}}$) are the electron creation (annihilation) operators with spin $\sigma$ at site $\mathbf{r}$ (here $\mathbf{r}$ is a vector with integer-valued components), $\eta$ runs from 1 to the dimension of the space $d$, $\hat{\eta}$ is a unit vector in a corresponding direction, and $C_{\eta\mathbf{r}} \equiv \psi_{\uparrow\mathbf{r}}\psi_{\downarrow\mathbf{r}+\hat{\eta}} - \psi_{\downarrow\mathbf{r}}\psi_{\uparrow\mathbf{r}+\hat{\eta}}$. The second term in (2.1) is just the superexchange interaction $J_\eta \left( \mathbf{S}_\mathbf{r} \cdot \mathbf{S}_{\mathbf{r}+\hat{\eta}} - \frac{1}{4} n_\mathbf{r} n_{\mathbf{r}+\hat{\eta}} \right)$, where $\mathbf{S}_\mathbf{r}$ is a spin operator and $n_\mathbf{r}$ is a number operator at site $\mathbf{r}$.

In (2.1) we have taken into account that high-temperature superconductors are all strongly anisotropic materials with direction-dependent parameters. To avoid the countless repetitions of the sign of summation over $\eta$, we will utilize an agreement (the generalization of the usual Einstein rule) that this sign will not be written explicitly in formulas where the index $\eta$ is used two or more times independently of its position. For example, summation over $\eta$ will be assumed in such expressions as $t_\eta^{-1} e^{ik_\eta} \cos k_\eta$. The geometrical meaning of this rule becomes clear in a long-wavelength limit, where all expressions of that type turn out to be scalars, created in the usual way from vectors and diagonalized tensors. This rule will be *not* applied in cases when the expression with repeating $\eta$-indices stands on one side of an equality, if the other side of the equality also has the index $\eta$, but used only once.

As was pointed out in the Introduction, in order to impose the single-occupancy constraint, the electron operators are usually represented as a product (1.1). We will see however, that it is more convenient, especially in working with diagrams, to use a different, although equivalent, representation

$$\psi^\dagger_{\sigma\mathbf{r}} = b^\dagger_{\sigma\mathbf{r}} f^\dagger_\mathbf{r}, \tag{2.2}$$

where $f^\dagger_\mathbf{r} = a_\mathbf{r}$. The constraint now has the form

$$b^\dagger_{\sigma\mathbf{r}} b_{\sigma\mathbf{r}} = f^\dagger_\mathbf{r} f_\mathbf{r}. \tag{2.3}$$

Due to the fermionic nature of the operators $f_\mathbf{r}$, the constraint (2.3) automatically restricts the possible number of SB and, of course, electrons at site $\mathbf{r}$ to 0 or 1.

The representation similar to (2.2) with the constraint (2.3) works equally well for the large-$U$ Hubbard model with more then one electron per site (i.e., for electron doping). There, the constrained subspace has one or two electrons on every site; the states with empty sites are excluded. We can then consider a doubly occupied site as corresponding to the absence of $b$ and $f$ fields, and a site with one electron as corresponding to the presence of one $f$ and one $b$ particle. An empty site would correspond here to the presence of two $f$ particles at one site, which is again impossible due to their fermionic nature. In each case, therefore, the presence of an $f$ particle means the presence of one and only one electron, so we will use for them the name *monon*.

The physical interpretation of the representation (2.2) is that in order to satisfy the condition of no doubly occupation, we can choose the basis many-electron wave functions of the constrained space in a form

$$\Psi_{\sigma_1,\ldots,\sigma_\mathcal{N}} (\mathbf{r}_1,\ldots,\mathbf{r}_\mathcal{N}) = \Psi^f (\mathbf{r}_1,\ldots,\mathbf{r}_\mathcal{N}) \Psi^b_{\sigma_1,\ldots,\sigma_\mathcal{N}} (\mathbf{r}_1,\ldots,\mathbf{r}_\mathcal{N}), \tag{2.4}$$

where $\Psi^f$ ($\Psi^b$) is an antisymmetric (symmetric) function of its arguments. We can then regard excitations $b$ and $f$, which correspond to wave functions $\Psi^b$ and $\Psi^f$, as the original elementary excitations in our theory. Their occupation numbers must satisfy the constraint (2.3), and their relation to electrons is given by (2.2). By (2.4), this relation is *local* in a sense, that both particles $b$ and $f$ occupy the same sites as electrons. This is in contrast to the *nonlocal* relation, corresponding to the usual representation (1.1), where spinons $b$ occupy the same sites as electrons, while holons $a$ occupy the remaining sites. It is interesting to note that the slave-boson formulation, which uses the same representation (1.1), but assigns fermionic statistics to operators $b_{\sigma\mathbf{r}}$ and bosonic statistics to operators $a_{\mathbf{r}}$, cannot be transformed to the monon language.

The substitution of (2.2) into (2.1) gives on the constrained space

$$H = \sum_{\mathbf{r}} \left[ -t_\eta \left( B_{\eta\mathbf{r}} F_{\eta\mathbf{r}} + B_{\eta\mathbf{r}}^\dagger F_{\eta\mathbf{r}}^\dagger \right) - \frac{J_\eta}{2} \mathcal{A}_{\eta\mathbf{r}}^\dagger \mathcal{A}_{\eta\mathbf{r}} \right], \qquad (2.5)$$

where $B_{\eta\mathbf{r}} \equiv b_{\sigma\mathbf{r}}^\dagger b_{\sigma\mathbf{r}+\hat{\eta}}$, $F_{\eta\mathbf{r}} \equiv f_{\mathbf{r}}^\dagger f_{\mathbf{r}+\hat{\eta}}$ and $\mathcal{A}_{\eta\mathbf{r}} \equiv b_{\uparrow\mathbf{r}} b_{\downarrow\mathbf{r}+\hat{\eta}} - b_{\downarrow\mathbf{r}} b_{\uparrow\mathbf{r}+\hat{\eta}}$. For the spinon's fields we will use Nambu's notation:

$$b_{\mathbf{r}} \equiv \begin{pmatrix} b_{\uparrow\mathbf{r}} \\ b_{\downarrow\mathbf{r}}^\dagger \end{pmatrix}, \quad b_{\mathbf{r}}^\dagger \equiv \left( b_{\uparrow\mathbf{r}}^\dagger, b_{\downarrow\mathbf{r}} \right) \qquad (2.6)$$

and Pauli matrices $\tau_j$, $j = 1, 2, 3$ and $\tau_\pm = \tau_1 \pm i\tau_2$; besides, we denote $\pi_\pm \equiv 1 \pm \tau_3$. For the operators $\mathcal{A}_{\eta\mathbf{r}}$, $B_{\eta\mathbf{r}}$, $F_{\eta\mathbf{r}}$ and their conjugates we will also use the general notation $M_{\alpha\eta\mathbf{r}}$, $\alpha = 1,\ldots,6$ with the identification

$$M_{\alpha\eta\mathbf{r}} \equiv \left( \mathcal{A}_{\eta\mathbf{r}}^\dagger, \mathcal{A}_{\eta\mathbf{r}}, B_{\eta\mathbf{r}}^\dagger, B_{\eta\mathbf{r}}, F_{\eta\mathbf{r}}^\dagger, F_{\eta\mathbf{r}} \right). \qquad (2.7)$$

We have then

$$\begin{aligned} M_{1,2\eta\mathbf{r}} &= \pm\frac{1}{2} \left( b_{\mathbf{r}}^\dagger \tau_\pm b_{\mathbf{r}+\hat{\eta}} - b_{\mathbf{r}+\hat{\eta}}^\dagger \tau_\pm b_{\mathbf{r}} \right), \\ M_{3,4\eta\mathbf{r}} &= \frac{1}{2} \left( b_{\mathbf{r}}^\dagger \pi_\mp b_{\mathbf{r}+\hat{\eta}} + b_{\mathbf{r}+\hat{\eta}}^\dagger \pi_\pm b_{\mathbf{r}} \right) \end{aligned} \qquad (2.8)$$

Besides operators (2.7), we also introduce the operator

$$M_{7\mathbf{r}} \equiv i \left( b_{\mathbf{r}}^\dagger b_{\mathbf{r}} - f_{\mathbf{r}}^\dagger f_{\mathbf{r}} \right). \qquad (2.9)$$

We will denote all $M$-operators as $M_{\alpha\eta\mathbf{r}}$, keeping in mind that $M_{7\mathbf{r}}$ does not have the index $\eta$. For the frequently used sums over $\alpha$ we will utilize the following notations: $\sum_\alpha = \sum_{\alpha=1}^7$, $\sum_{\alpha_{b,f}} = \sum_{\alpha \in \mathcal{N}_{b,f}}$, and $\sum_{\widetilde{\alpha}_{b,f}} = \sum_{\alpha \in \widetilde{\mathcal{N}}_{b,f}}$, where $\mathcal{N}_b = \{1,\ldots,4\}$, $\mathcal{N}_f = \{5,6\}$, $\widetilde{\mathcal{N}}_b = \{1,\ldots,4,7\}$, and $\widetilde{\mathcal{N}}_f = \{5,6,7\}$.

The constraint has the form

$$b_{\mathbf{r}}^\dagger b_{\mathbf{r}} = f_{\mathbf{r}}^\dagger f_{\mathbf{r}}, \qquad (2.10)$$

or $M_{7\mathbf{r}} = 0$. In all formulas which include the product of SB operators in Nambu's notation, the normal ordering of initial operators $b_{\sigma\mathbf{r}}$, $b_{\sigma\mathbf{r}}^\dagger$ is assumed.

A convenient way to work with Hamiltonian (2.5) and the constraint is the $1/N$ expansion. We generalize our model to $N \geq 1$ colors by setting $t_\eta \to t_\eta/N$, $J_\eta \to J_\eta/N$, $b_{\mathbf{r}}(f_{\mathbf{r}}) \to b_{\mathbf{r}}^\alpha(f_{\mathbf{r}}^\alpha)$, $\alpha = 1,\ldots,N$. The product of boson or fermion fields within each composite operator $M_{\alpha\eta\mathbf{r}}$ is now understood as a scalar product with respect to the color index; the constraint (2.10) is generalized to $b_{\mathbf{r}}^{\alpha\dagger} b_{\mathbf{r}}^\alpha = f_{\mathbf{r}}^{\alpha\dagger} f_{\mathbf{r}}^\alpha$. In the following, we will not write the color indices explicitly, since it will be always clear what kind of scalar product with respect to them is used.

To impose the constraint, we add to the Hamiltonian the new term $H_\epsilon \equiv -(2N\epsilon)^{-1} \sum_{\mathbf{r}} M_{7\mathbf{r}}^2$. As was demonstrated in [9], keeping $\epsilon$ finite helps to control infrared divergences, which must cancel in each order of the $1/N$ expansion separately. The limit $\epsilon \to 0$ must be taken at the end of the calculations. We now introduce the $7 \times 7$ matrix $h_\eta$ with nonzero matrix elements $h_{\eta 12} = h_{\eta 21} = J_\eta/2$, $h_{\eta 35} = h_{\eta 53} = h_{\eta 46} = h_{\eta 64} = t_\eta$, $h_{\eta 77} = 1/\epsilon$ and write the partition function as an imaginary-time path integral:

$$\begin{aligned} Z &= \int \mathcal{D}(b^* b f^* f) \exp\left[ -\int dx \left( b_x^* \tau_3 \dot{b}_x + f_x^* \dot{f}_x \right.\right.\\ &\left.\left. - \frac{1}{2N} \sum_{\alpha\beta} M_{\alpha\eta x} h_{\eta\alpha\beta} M_{\beta\eta x} - \mu n_x \right) \right]. \end{aligned} \qquad (2.11)$$

Here the chemical potential $\mu$ fixes the electron density and we have used the notations $n_x \equiv (1/2)(b_x^* b_x + f_x^* f_x)$, $x \equiv (\tau, \mathbf{r})$, $\int dx \equiv \int_0^\beta d\tau \sum_{\mathbf{r}}$. We now introduce the real auxiliary field (AF) $\phi_{7x}$ and the set of complex AF $\phi_{\alpha \eta x}$, $\alpha = 1, \ldots, 6$, which are divided into three pairs of conjugated fields according to

$$\phi_{2\eta x} = \phi_{1\eta x}^*, \quad \phi_{5\eta x} = \phi_{3\eta x}^*, \quad \phi_{6\eta x} = \phi_{4\eta x}^*. \tag{2.12}$$

We will use for all AF the notation $\phi_{\alpha \eta x}$, with the understanding that $\phi_{7x}$ does not have the index $\eta$. Then the interaction terms in (2.11) may be decoupled, yielding

$$Z = \int \mathcal{D}(b^* b f^* f \phi) \exp\left[-\int dx \left(b_x^* \tau_3 \dot{b}_x + f_x^* \dot{f}_x + \frac{1}{2} \sum_{\alpha \beta} \phi_{\alpha \eta x} \Pi^{(0)}_{\eta \alpha \beta} \phi_{\beta \eta x} \right.\right. \\ \left.\left. - \sum_\alpha \phi_{\alpha \eta x} M_{\alpha \eta x} - \mu n_x \right)\right], \tag{2.13}$$

where the bare polarization matrix of AF $\Pi^{(0)}_\eta$ is given by

$$\Pi^{(0)}_{\eta \alpha \beta} \equiv N\left(h_\eta^{-1}\right)_{\alpha \beta} = \begin{cases} 0 & \text{for} \quad h_{\eta \alpha \beta} = 0 \\ \dfrac{N}{h_{\eta \alpha \beta}} & \text{for} \quad h_{\eta \alpha \beta} \neq 0 \end{cases} \tag{2.14}$$

Equation (2.13) is invariant with respect to the local gauge transformation, induced by the representation (2.2) and the constraint (2.3). To make this gauge invariance explicit, we use the following parametrization of AF $\phi_{\alpha \eta x}$, $\alpha = 1, \ldots, 6$:

$$\phi_{\alpha \eta x} = |\phi_{\alpha \eta x}| \exp\left\{ il \left[\tilde{g}_\alpha \tilde{a}_{\eta x} + g_\alpha (a_{\eta x} + j_\alpha \theta_{\eta x})\right]\right\}, \tag{2.15}$$

where $l$ is a lattice constant and $\tilde{g}_\alpha$, $g_\alpha$, $j_\alpha$, $\alpha = 1, \ldots, 7$ are given by

$$\begin{aligned}
\tilde{g}_\alpha &= (\ -1,\ \ 1,\ \ 0,\ \ 0,\ \ \ 0,\ \ \ 0,\ \ 0\ ), \\
g_\alpha &= (\ \ 0,\ \ 0,\ \ 1,\ -1,\ -1,\ \ 1,\ \ 0\ ), \\
j_\alpha &= (\ \ 0,\ \ 0,\ \ 1,\ -1,\ \ \ 1,\ -1,\ \ 0\ ).
\end{aligned} \tag{2.16}$$

We have then

$$\tilde{a}_{\eta x} = \frac{1}{2il} \sum_\alpha \tilde{g}_\alpha \ln \phi_{\alpha \eta x}, \tag{2.17}$$

$$a_{\eta x} = \frac{1}{4il} \sum_\alpha g_\alpha \ln \phi_{\alpha \eta x}. \tag{2.18}$$

Now in the limit $\epsilon \to 0$ Eq. (2.13) is invariant with respect to the local gauge transformation

$$b_{\sigma x} \longrightarrow e^{i\chi_x} b_{\sigma x}, \quad f_x \longrightarrow e^{-i\chi_x} f_x, \tag{2.19}$$

$$a_{0x} \longrightarrow a_{0x} + \partial_0 \chi_x, \tag{2.20}$$

$$\tilde{a}_{\eta x} \longrightarrow \tilde{a}_{\eta x} + \tilde{\partial}_\eta^R \chi_x, \quad a_{\eta x} \longrightarrow a_{\eta x} + \partial_\eta^R \chi_x. \tag{2.21}$$

Here $a_{0x} \equiv \phi_{7x}$, $\partial_0 \equiv \partial/(\partial \tau)$, $\partial_\eta^R$ is a "right" lattice derivative, $\partial_\eta^R \chi_x \equiv l^{-1}(\chi_{x+\hat{\eta}} - \chi_x)$, where we define $x \pm \hat{\eta} \equiv (\tau, \mathbf{r} \pm \hat{\eta})$ and $\tilde{\partial}_\eta^R$ is a "staggered" right lattice derivative, defined as $\tilde{\partial}_\eta^R \equiv \xi_{\mathbf{r}} \partial_\eta^R \xi_{\mathbf{r}}$. Here $\xi_{\mathbf{r}} \equiv \exp(i\vec{\pi}\mathbf{r})$, where $\vec{\pi}$ is a vector which has all components equal to $\pi$; obviously, $\tilde{\partial}_\eta^R \chi_x = -l^{-1}(\chi_{x+\hat{\eta}} + \chi_x)$. We see, that the gauge invariance in this problem has a nonstandard form, being characterized by the presence of two gauge fields: "staggered" $\tilde{a}_{\eta x}$ and "uniform" $a_{\eta x}$.

## II.2 Diagram technique and self-consistency equations

### II.2.1 Real space

Due to the gauge invariance, Eq. (2.13) contains redundant integration, and in order to develop the diagram technique we must fix the gauge. We will do this by adding to the Lagrangian in (2.13) the gauge-fixing term (GFT) $\mathcal{L}_g(x)$. The explicit form of the GFT will be presented later. After fixing the gauge the original local symmetry is lost, so the remaining symmetries are global and may be spontaneously broken. Due to this spontaneous symmetry breaking, the spinon fields $b_x$ as well as AF $\phi_{\alpha\eta x}$ and composite fields $M_{\alpha\eta x}$, $\alpha = 1, \ldots, 6$ may obtain nonzero expectation values. As was discussed in the Introduction, although these fields are not gauge-invariant, with a fixed gauge this is possible and does not contradict Elitzur's theorem [7]. We also assume the possibility of having nonzero expectation values for the field $\phi_{7x}$ and the composite operator $M_{7x}$ (see the corresponding discussion below, after Eq. (2.27)). Consequently, we make in (2.13) the shifts $b_x \to v_{\mathbf{r}} + b_x$ and $\phi_{\alpha\eta x} \to \phi_{\alpha\eta} + \phi_{\alpha\eta x}$, where $v_{\mathbf{r}}$ and $\phi_{\alpha\eta}$ are the expectation values of the corresponding fields while $b_x$ and $\phi_{\alpha\eta x}$ are now their fluctuating parts. Here parameter $v_{\mathbf{r}}$ is time and color independent and for the general discussion in this paper we only consider a simple case of space and time independent parameters $\phi_{\alpha\eta}$. The choice of $v_{\mathbf{r}}$ and $\phi_{\alpha\eta}$ will be specified later. It will be clear then, that these parameters are determined on the MF level and do not depend on the gauge choice. Therefore, it is legitimate to choose a gauge condition, which itself depends on these parameters. In other words, we first determine the phase of the model, and then choose a gauge, which is appropriate in this phase. At this moment we will leave the GFT in the representation (2.22) below in an unspecified form.

We now arrive at the final path-integral representation for the partition function

$$\begin{aligned}
Z &= \int \mathcal{D}(b^*b f^*f \phi) \exp\left[-\int \mathcal{L}(x)dx\right], \\
\mathcal{L}(x) &= \mathcal{L}_{MF}(x) + \mathcal{L}_{lin}(x) + \mathcal{L}_{AF}^{(0)}(x) + \mathcal{L}_g(x) + \mathcal{L}_{int}(x) + \text{const}, \\
\mathcal{L}_{MF}(x) &= b_x^* \tau_3 \dot{b}_x + f_x^* \dot{f}_x - \sum_\alpha \phi_{\alpha\eta} M_{\alpha\eta x} - \mu n_x, \\
\mathcal{L}_{lin}(x) &= -\sum_{\alpha_b} \phi_{\alpha\eta} \widetilde{M}_{\alpha\eta x} - \mu_b(b_x^* v_{\mathbf{r}} + v_{\mathbf{r}}^\dagger b_x), \\
\mathcal{L}_{AF}^{(0)}(x) &= \frac{1}{2} \sum_{\alpha\beta} \phi_{\alpha\eta x} \Pi^{(0)}_{\eta\alpha\beta} \phi_{\beta\eta x}, \\
\mathcal{L}_{int}(x) &= -\sum_\alpha \phi_{\alpha\eta x} M'_{\alpha\eta x},
\end{aligned} \quad (2.22)$$

where

$$\begin{aligned}
\widetilde{M}_{\alpha\eta x} &\equiv M_{\alpha\eta x}\Big|_{b_x \to v_{\mathbf{r}} + b_x} - M_{\alpha\eta x}\Big|_{b_x \to v_{\mathbf{r}}} - M_{\alpha\eta x}, \\
M'_{\alpha\eta x} &\equiv M_{\alpha\eta x}\Big|_{b_x \to v_{\mathbf{r}} + b_x} - N M_{\alpha\eta}, \\
M_{\alpha\eta} &\equiv \frac{1}{N} \sum_\beta \Pi^{(0)}_{\eta\alpha\beta} \phi_{\beta\eta},
\end{aligned} \quad (2.23)$$

and we denote $\mu_{b,f} \equiv \pm i\phi_7 + \mu/2$. $\mathcal{L}_{lin}(x)$ includes terms, that are linear in spinon fields.

The representation (2.22) generates the diagram technique in our theory, with $\mathcal{L}_{MF}$ regarded as a free Lagrangian of $b$ and $f$ fields, which determines their bare propagators. On the diagrams, we will represent the propagators of fields $b$ and $f$ by solid lines, and the propagators of AF (which will be determined later) by wavy lines. The factors $v_{\mathbf{r}}$, $v_{\mathbf{r}}^\dagger$ will be represented by short zigzag tails. We note, that if we express $\phi_{\alpha\eta}$ in $\mathcal{L}_{MF}$ as

$$\phi_{\alpha\eta} = \sum_\beta h_{\eta\alpha\beta} M_{\beta\eta} \quad (2.24)$$

then $\mathcal{L}_{MF}$ will correspond to the MF decomposition of the Hamiltonian (2.5).

In the following, we will use for the arbitrary functional of the fields $\mathcal{F}$ the notation

$$\langle \mathcal{F} \rangle \equiv Z^{-1} \int \mathcal{D}(b^*bf^*f\phi)\, \mathcal{F}\, e^{-\int \mathcal{L}(x)dx}. \tag{2.25}$$

The notation $\langle \mathcal{F} \rangle_{MF}$ will have the same meaning, with the changes $\mathcal{L} \to \mathcal{L}_{MF}$ and $Z \to Z_{MF}$, where $Z_{MF}$ is the partition function of the MFT with the Lagrangian $\mathcal{L}_{MF}$.

To determine the unknown parameters $\mu$, $v_{\mathbf{r}}$ and $\phi_{\alpha\eta}$ (or $M_{\alpha\eta}$) we will use the following conditions. First, if $\delta$ is a density of holes, then we must have

$$\langle n_x \rangle = N(1-\delta). \tag{2.26}$$

Second, we require that the part of the action linear in $b_x$, $b_x^*$, i.e., $\mathcal{L}_{lin}(x)$, must vanish; it will become clear below that this condition yields $\lim_{N\to\infty}\langle b_x \rangle = 0$. We will return to the explicit form of this condition later. Finally, we require the vanishing of the "bare tadpole" subdiagrams of Fig. 1(a), so that the diagrams with these subdiagrams must be excluded. This condition is $\langle M'_{\alpha\eta x}\rangle_{MF} = 0$, or

$$\langle M_{\alpha\eta x}\Big|_{b_x \to v_{\mathbf{r}}+b_x} \rangle_{MF} = N M_{\alpha\eta}. \tag{2.27}$$

Thus the MF expectation values of operators $M_{\alpha\eta x}$ per color must be equal to the values $M_{\alpha\eta}$, introduced in

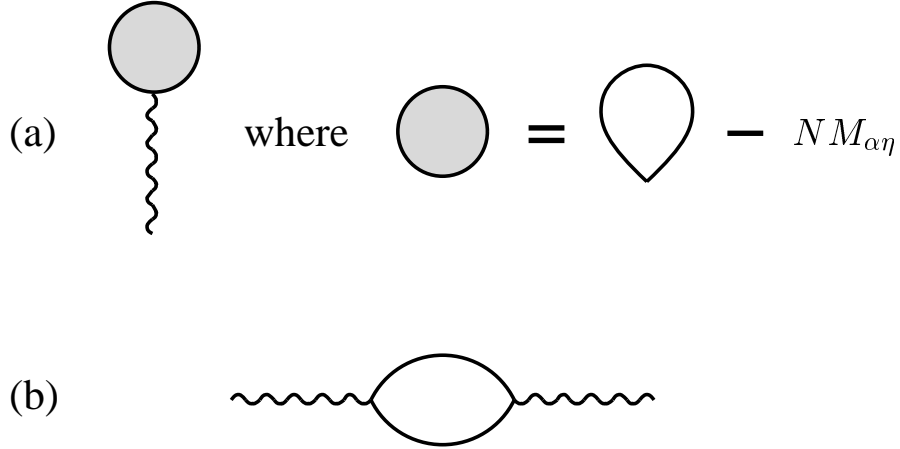

FIG. 1. Forbidden (sub)diagrams. (a) "Bare tadpoles." The shaded circle represents $\langle M'_{\alpha\eta x}\rangle_{MF} = \langle M_{\alpha\eta x}\rangle_{MF} - NM_{\alpha\eta}$. Shown is the case $v_{\mathbf{r}} = 0$; when $v_{\mathbf{r}} \neq 0$ and $\alpha \in \widetilde{\mathcal{N}}_b$, the obvious term with two zigzag tails must be added to the shaded circle. (b) RPA bubble, inserted to the AF line.

Eq. (2.23). For $\alpha = 7$ we note, that with our softened way of introducing the constraint, the operator $M_{7x}$ is allowed to have a small expectation value, controlled by parameter $\epsilon$. In order to obtain the finite results in the limit $\epsilon \to 0$, we take $\phi_7$ to be independent of $\epsilon$, so that $M_7 = \epsilon\phi_7$ vanishes as $\epsilon$ for $\epsilon \to 0$. The value of $\phi_7$ itself is then determined by Eq. (2.27) with $\alpha = 7$, which in the limit $\epsilon \to 0$ gives

$$|v_{\mathbf{r}}|^2 + \langle b_x^* b_x \rangle_{MF} = \langle f_x^* f_x \rangle_{MF}, \tag{2.28}$$

where we used the equality $\langle b_x \rangle_{MF} = 0$. We note that the parameters $M_{\alpha\eta}$ are determined on the MF level by equations (2.26) – (2.28); the exact expectation values $\langle M_{\alpha\eta x}\rangle$ might be, of course, different from the corresponding MF expectation values, used in these equations.

### II.2.2 Momentum space

We will now derive explicit expressions for the bare Green's functions of spinons and monons and their interaction vertices with AF. We denote $k \equiv (\omega, \mathbf{k})$, $kx \equiv \omega\tau - \mathbf{k}\mathbf{r}$ and $\sum_k \equiv (\beta V)^{-1} \sum_{\omega, \mathbf{k}} = T \sum_\omega \int_\Omega \frac{d^d k}{(2\pi)^d}$, where $\omega$ is a Matsubara frequency, $\mathbf{k}$ is a lattice momentum, $V$ is the number of lattice sites, and $\Omega$ refers to the first Brillouin zone. The Fourier transformation is defined according to

$$\psi_k = \int dx\, \psi_x e^{ikx}, \qquad \psi_x = \sum_k \psi_k e^{-ikx} \tag{2.29}$$

for an arbitrary value $\psi_x$, while the value $\psi_x^\dagger$ is transformed according to the conjugate definition. For the Nambu spinors (2.6) this gives

$$b_k = \begin{pmatrix} b_{\uparrow k} \\ b_{\downarrow -k}^\dagger \end{pmatrix}, \qquad b_k^\dagger = \begin{pmatrix} b_{\uparrow k}^\dagger, b_{\downarrow -k} \end{pmatrix}. \tag{2.30}$$

The Fourier components of the operators $M$ are

$$M_{\alpha\eta q} = \begin{cases} e^{i\frac{q_\eta}{2}} \sum_k b_{k-\frac{q}{2}}^\dagger m_{\alpha\eta\mathbf{k}} b_{k+\frac{q}{2}}, & \alpha \in \mathcal{N}_b \\ e^{i\frac{q_\eta}{2}} \sum_k f_{k-\frac{q}{2}}^\dagger m_{\alpha\eta\mathbf{k}} f_{k+\frac{q}{2}}, & \alpha \in \mathcal{N}_f \end{cases} \tag{2.31}$$

$$M_{7q} = i\sum_k \left( b_{k-\frac{q}{2}}^\dagger b_{k+\frac{q}{2}} - f_{k-\frac{q}{2}}^\dagger f_{k+\frac{q}{2}} \right),$$

where

$$\begin{aligned} m_{1,2\eta\mathbf{k}} &= \pm i\tau_\pm \sin k_\eta, \\ m_{3,4\eta\mathbf{k}} &= \frac{1}{2}\left(\pi_\mp e^{ik_\eta} + \pi_\pm e^{-ik_\eta}\right), \\ m_{5,6\eta\mathbf{k}} &= e^{\mp ik_\eta}. \end{aligned} \tag{2.32}$$

For the MF action in Eq. (2.22) we have

$$\begin{aligned} \mathcal{S}_{MF} = \sum_k \Bigg[ & b_k^* \left( -i\omega\tau_3 - \mu_b - \sum_{\alpha_b} \phi_{\alpha\eta} m_{\alpha\eta\mathbf{k}} \right) b_k \\ & + f_k^* \left( -i\omega - \mu_f - \sum_{\alpha_f} \phi_{\alpha\eta} m_{\alpha\eta\mathbf{k}} \right) f_k \Bigg]. \end{aligned} \tag{2.33}$$

The bare Green's function of spinons, $G_x \equiv \langle b_x b_0^* \rangle_{MF}$, has the Fourier transformation

$$G_k = \frac{1}{\beta V} \langle b_k b_k^* \rangle_{MF} = \frac{1}{\beta V} \begin{pmatrix} \langle b_{\uparrow k} b_{\uparrow k}^* \rangle_{MF} & \langle b_{\uparrow k} b_{\downarrow -k} \rangle_{MF} \\ \langle b_{\downarrow -k}^* b_{\uparrow k}^* \rangle_{MF} & \langle b_{\downarrow -k}^* b_{\downarrow -k} \rangle_{MF} \end{pmatrix}. \tag{2.34}$$

From (2.33) we have for it

$$G_k = \left( -i\omega\tau_3 - \mu_b - \sum_{\alpha_b} \phi_{\alpha\eta} m_{\alpha\eta\mathbf{k}} \right)^{-1}. \tag{2.35}$$

The bare Green's function of monons $g_x \equiv \langle f_x f_0^* \rangle_{MF}$ has the Fourier transformation

$$g_k = \frac{1}{\beta V} \langle f_k f_k^* \rangle_{MF} = \left( -i\omega - \mu_f - \sum_{\alpha_f} \phi_{\alpha\eta} m_{\alpha\eta\mathbf{k}} \right)^{-1}. \tag{2.36}$$

In terms of these Green's functions, the stationarity equations (2.27), (2.28) are

$$\sum_k \text{Tr}\,(m_{\alpha\eta\mathbf{k}} G_k) + \sum_{\mathbf{k}} v_{\mathbf{k}}^\dagger m_{\alpha\eta\mathbf{k}} v_{\mathbf{k}} = M_{\alpha\eta}, \qquad \alpha \in \mathcal{N}_b \tag{2.37}$$

$$\sum_k m_{\alpha\eta\mathbf{k}} g_k = -M_{\alpha\eta}, \qquad \alpha \in \mathcal{N}_f \tag{2.38}$$

$$\sum_k \text{Tr}\,G_k + \sum_{\mathbf{k}} v_{\mathbf{k}}^\dagger v_{\mathbf{k}} = -\sum_k g_k, \tag{2.39}$$

where Tr traces over Nambu indices, the minus sign is due to the fermionic statistics of monons and we have written the Fourier transformation for the time-independent parameter $v_{\mathbf{r}}$ as $v_{\mathbf{r}} = \sum_{\mathbf{k}} v_{\mathbf{k}} \exp(i\mathbf{k}\mathbf{r})$ (here the sum over $\mathbf{k}$ does not include the factor of $1/V$). The factor $e^{i\omega\delta}$, $\delta = +0$, is assumed to be inserted under the sums over $k$ in these equations [10]. Now the obvious identity $\sum_k \frac{\partial}{\partial k_\eta} \ln g_k^{-1} = 0$ and Eqs. (2.32), (2.36), (2.38) yield

$$it_\eta (M_{4\eta} M_{6\eta} - M_{3\eta} M_{5\eta}) = 0 \tag{2.40}$$

or $\text{Im}(M_{4\eta} M_{6\eta}) = 0$, so the phases of $M_{4\eta}$ and $M_{6\eta}$ are opposite. Consequently, in the states with zero values of the parameters $\mathcal{A}_\eta \equiv M_{2\eta}$ by the gauge transformation $a_{\eta x} \to a_{\eta x} + \delta a_\eta$ with a constant $\delta a_\eta$ we can make all parameters $M_{\alpha\eta}$, $\alpha = 3, \ldots, 6$ real and positive. For simplicity, we assume that they are also real and positive in the case of nonzero $\mathcal{A}_\eta$. We will use for these parameters the notations $B_\eta \equiv M_{b\eta} \equiv M_{3\eta} = M_{4\eta}$ and $F_\eta \equiv M_{f\eta} \equiv M_{5\eta} = M_{6\eta}$. Using Eqs. (2.38) and (2.43) below, one can easily see that $F_\eta$ is different from zero if and only if $B_\eta$ is different from zero [11].

Now the spinon Green's function (2.35) may be written in the explicit form:

$$\begin{aligned} G_k &= \left[ -i\omega\tau_3 - \mu_b - 2t_\eta F_\eta \cos k_\eta + J_\eta (\mathcal{A}'_\eta \tau_2 + \mathcal{A}''_\eta \tau_1) \sin k_\eta \right]^{-1} \\ &= \frac{i\omega\tau_3 - \mu_b - 2t_\eta F_\eta \cos k_\eta - J_\eta (\mathcal{A}'_\eta \tau_2 + \mathcal{A}''_\eta \tau_1) \sin k_\eta}{\omega^2 + (\mu_b + 2t_\eta F_\eta \cos k_\eta)^2 - |J_\eta \mathcal{A}_\eta \sin k_\eta|^2}, \end{aligned} \tag{2.41}$$

where $\mathcal{A}'_\eta$ and $\mathcal{A}''_\eta$ are the real and imaginary parts of $\mathcal{A}_\eta$, respectively. The dispersion law of spinons is given by

$$\omega_{\mathbf{k}} = \sqrt{(\mu_b + 2t_\eta F_\eta \cos k_\eta)^2 - |J_\eta \mathcal{A}_\eta \sin k_\eta|^2}. \tag{2.42}$$

The monon Green's function (2.36) may be written explicitly as

$$g_k = (-i\omega - \mu_f - 2t_\eta B_\eta \cos k_\eta)^{-1} \equiv (-i\omega + \epsilon_{\mathbf{k}})^{-1}, \tag{2.43}$$

where the quantity $\epsilon_{\mathbf{k}}$ defined by this equation gives the monon dispersion law.

The parameter $v_{\mathbf{r}}$ is determined by requiring the linear part of the action $\mathcal{S}_{lin} \equiv \int \mathcal{L}_{lin}(x)\,dx$ to vanish. We have

$$\mathcal{S}_{lin} = \sum_{\mathbf{k}} (b_{\mathbf{k}}^* G_{\mathbf{k}}^{-1} v_{\mathbf{k}} + v_{\mathbf{k}}^\dagger G_{\mathbf{k}}^{-1} b_{\mathbf{k}}), \tag{2.44}$$

where we have denoted $G_{\mathbf{k}}^{-1} \equiv G_k^{-1}|_{k=(0,\mathbf{k})}$, and similarly for $b_{\mathbf{k}}$. Now $v_{\mathbf{k}}$ is to be found from the equation $G_{\mathbf{k}}^{-1} v_{\mathbf{k}} = 0$, which has a solution only for $\mathbf{k} = \mathbf{k}_c$ such that $\omega_{\mathbf{k}_c} = 0$. Therefore, for $\mathbf{k} \neq \mathbf{k}_c$, $v_{\mathbf{k}} = 0$, while for $\mathbf{k} = \mathbf{k}_c$ we obtain

$$v_{\mathbf{k}_c} = \alpha_{\mathbf{k}_c} \begin{pmatrix} -iJ_\eta \mathcal{A}_\eta \sin k_{c\eta} \\ \mu_b + 2t_\eta F_\eta \cos k_{c\eta} \end{pmatrix}, \tag{2.45}$$

where $\alpha_{\mathbf{k}_c}$ is a coefficient of proportionality. Evidently, if $\omega_{\mathbf{k}_c} = 0$ then also $\omega_{-\mathbf{k}_c} = 0$; the coefficients $\alpha_{\pm \mathbf{k}_c}$ along with the other parameters are to be obtained from the self-consistency equations (2.37) – (2.39). We note that these equations determine only the sum $|\alpha_{\mathbf{k}_c}|^2 + |\alpha_{-\mathbf{k}_c}|^2$. Therefore, with this sum given, the phases of $\alpha_{\pm \mathbf{k}_c}$ and their relative absolute values may be chosen arbitrarily, which reflects the existence of different possible ways of breaking the symmetry.

Instead of performing a shift of spinon fields, one can use a different, and often more convenient way to describe the effect of spinon condensation [12, 13]. Namely, since we consider a large, but finite lattice, the expectation value of $b_x$ may be different from zero only in the presence of the infinitesimally small ordering external field, which points the direction of the symmetry breaking. In the absence of such field, the expectation value of $b_x$ will vanish, and the ordering will manifest itself in the appearance of the abnormally large (of order $\beta V$) values of the boson Green's function at momentums $k_c \equiv (0, \mathbf{k}_c)$. Indeed, in this case the MF order parameters are such that $\omega_{\mathbf{k}_c}$ is not exactly equal to zero, but has (for nonzero temperature) a small but finite value of order $(\beta V)^{-1/2}$ [12]. From (2.41) we have then $G_{k_c} = \beta V v_{\mathbf{k}_c} v^\dagger_{\mathbf{k}_c}$, where $v_{\mathbf{k}_c}$ is given by (2.45) with $\alpha_{\mathbf{k}_c} = [-\beta V \omega^2_{\mathbf{k}_c}(\mu_b + 2t_\eta F_\eta \cos k_{c\eta})]^{-1/2}$. In this way, the theory of the phase with $v_\mathbf{r} \neq 0$ may be presented in the same form as that for $v_\mathbf{r} = 0$, and one only has to carefully separate out the terms with $k = k_c$ in the sums over $k$. Formally, this separation is achieved by the substitution

$$G_k \longrightarrow G_k + \beta V \sum_{\mathbf{k}_c} v_{\mathbf{k}_c} v^\dagger_{\mathbf{k}_c} \delta_{k k_c}. \tag{2.46}$$

Finally, in the Fourier representation, the interaction action in (2.22) is

$$\mathcal{S}_{int} = -\sum_q \sum_\alpha \phi_{\alpha\eta - q} M'_{\alpha\eta q}. \tag{2.47}$$

Therefore, the AF–monon and AF–spinon interaction vertices $U_{\alpha\eta}(k, q)$ have the form

$$U_{\alpha\eta}(k, q) = e^{i\frac{q_\eta}{2}} m_{\alpha\eta \mathbf{k}}, \tag{2.48}$$

where for $\alpha = 7$ the exponential factor in this formula must be omitted, and by $m_{7\mathbf{k}}$ we mean $m^{b,f}_{7\mathbf{k}} \equiv \pm i$, depending on whether AF $\phi_{7q}$ interacts with a spinon or a monon.

## II.3  Gauge fixing and the AF sector

Here we want to discuss the AF sector of the diagram technique. First we need to choose a GFT. We define a "left" lattice derivative $\partial^L_\eta$, and its staggered counterpart $\tilde{\partial}^L_\eta$, by the relations $\partial^L_\eta \psi_x \equiv l^{-1}(\psi_x - \psi_{x-\hat{\eta}})$ and $\tilde{\partial}^L_\eta \equiv \xi_\mathbf{r} \partial^L_\eta \xi_\mathbf{r}$, so that $\tilde{\partial}^L_\eta \psi_x = l^{-1}(\psi_x + \psi_{x-\hat{\eta}})$. We then use the GFT

$$\begin{aligned} \mathcal{L}_g(x) &= \frac{N}{2\zeta} L^2_x, \\ L_x &= \frac{1}{8} J^2_\eta |\mathcal{A}_\eta|^2 \tilde{\partial}^L_\eta (\tilde{a}_{\eta x} - \tilde{a}_\eta) + t^2_\eta B_\eta F_\eta \partial^L_\eta (a_{\eta x} - a_\eta), \end{aligned} \tag{2.49}$$

where $\zeta$ is an arbitrary positive number and $\tilde{a}_\eta, a_\eta$ are the gauge fields at the saddle point:

$$\begin{aligned} \tilde{a}_\eta &= \frac{1}{2il} \sum_\alpha \tilde{g}_\alpha \ln \phi_{\alpha\eta}, \\ a_\eta &= \frac{1}{4il} \sum_\alpha g_\alpha \ln \phi_{\alpha\eta}. \end{aligned} \tag{2.50}$$

Evidently, in the case when parameters $M_{\alpha\eta}$, $\alpha = 3, \ldots, 6$ are real, $a_\eta$ is equal to zero. The reason for the choice of the GFT in the form (2.49) will become clear in the next section. Although $\tilde{a}_{\eta x}$ and $a_{\eta x}$ are logarithmic functions of the original AF, we will see below that in the framework of the $1/N$ expansion one can develop a diagram technique, using the GFT in the form (2.49). It may be easily seen that the Faddeev-Popov determinant, corresponding to the GFT (2.49), is independent of AF and hence may be ignored.

The GFT (2.49) really fixes the gauge only in the phases, where for every direction $\eta$ at least one of the quantities $\mathcal{A}_\eta$ and $B_\eta F_\eta$ is different from zero. This condition is not satisfied in the symmetric phase, which has vanishing $\mathcal{A}_\eta$, $B_\eta$, and $F_\eta$ for all $\eta$; however, in this work we will not be interested in this phase. Also, this condition is violated in the physically important phases with $\mathcal{A}_{\eta_0} = B_{\eta_0} = F_{\eta_0} = 0$ for only one direction $\eta_0$. In this case we will consider the problem as two-dimensional, neglecting the existence of the direction $\eta_0$; in the plane, orthogonal to $\eta_0$, our condition is satisfied, and we can use the GFT (2.49).

We note that the GFT (2.49) does not eliminate the gauge freedom completely: there still remains the residual symmetry with respect to the gauge transformations, which depend only on time and are space-independent. However, for nonzero $\epsilon$ our action is not invariant with respect to these transformations. Therefore, the corresponding divergences will produce contributions proportional to $1/\epsilon$; as was already pointed out, these contributions must cancel each other in every order of the $1/N$ expansion (see Ref. [9], where such cancellation was demonstrated explicitly in the first order in $1/N$). A different way to eliminate this residual gauge freedom will be discussed at the end of the next section.

We now want to learn how to work with the GFT (2.49). For that we note that $\tilde{a}_{\eta x}$ contributes to (2.49) only in the phases with $\phi_{1,2\eta} \neq 0$, and $a_{\eta x}$ – only in the phases with $\phi_{\alpha\eta} \neq 0$, $\alpha = 3, \ldots, 6$. We can then use the fact that the parametrization (2.15), which is not invertible at $\phi_{\alpha\eta x} = 0$, is invertible about the point $\phi_{\alpha\eta x} = \phi_{\alpha\eta}$. The inversion is given by the expansion of (2.17) – (2.18) around $\phi_{\alpha\eta}$ in powers of $\phi_{\alpha\eta x}$ as

$$\begin{aligned}
\tilde{a}_{\eta x} &= \frac{1}{2il} \sum_\alpha \tilde{g}_\alpha \ln(\phi_{\alpha\eta} + \phi_{\alpha\eta x}) = \tilde{a}_\eta + \tilde{\phi}_{\eta x} + \tilde{\phi}'_{\eta x}, \\
a_{\eta x} &= \frac{1}{4il} \sum_\alpha g_\alpha \ln(\phi_{\alpha\eta} + \phi_{\alpha\eta x}) = a_\eta + \phi_{\eta x} + \phi'_{\eta x},
\end{aligned} \quad (2.51)$$

where

$$\begin{aligned}
\tilde{\phi}_{\eta x} &= \frac{1}{2il} \sum_\alpha \tilde{g}_\alpha \frac{\phi_{\alpha\eta x}}{\phi_{\alpha\eta}}, \\
\phi_{\eta x} &= \frac{1}{4il} \sum_\alpha g_\alpha \frac{\phi_{\alpha\eta x}}{\phi_{\alpha\eta}},
\end{aligned} \quad (2.52)$$

and $\tilde{\phi}'_{\eta x}$, $\phi'_{\eta x}$ include all the terms of second and higher powers in AF, coming from the expansion of the logarithms in (2.51).

We have now for the GFT $\mathcal{L}_g(x) = \mathcal{L}_g^{(0)}(x) + \mathcal{L}'_g(x)$, where

$$\begin{aligned}
\mathcal{L}_g^{(0)}(x) &= \frac{N}{2\zeta} L_x^{(0)2}, \\
L_x^{(0)} &= \frac{1}{8} J_\eta^2 |\mathcal{A}_\eta|^2 \tilde{\partial}_\eta^L \tilde{\phi}_{\eta x} + t_\eta^2 B_\eta F_\eta \partial_\eta^L \phi_{\eta x},
\end{aligned} \quad (2.53)$$

and $\mathcal{L}'_g(x)$ includes the contribution of $\tilde{\phi}'_{\eta x}$ and $\phi'_{\eta x}$. We will regard $\mathcal{L}'_g(x)$ as describing the self-interaction of AF. Although it has an infinite number of terms, it will be clear that only a finite number of them contribute to every order of the $1/N$ expansion.

The effective free action of AF (the sum of all terms quadratic in AF and proportional to $N$) is given now by the bare part in (2.22), by the quadratic part of GFT and by the contribution of the random-phase approximation (RPA) polarization bubbles:

$$\begin{aligned}
\mathcal{S}_{eff} &= \int \left[ \mathcal{L}_{AF}^{(0)}(x) + \mathcal{L}_g^{(0)}(x) \right] dx \\
&\quad - \frac{1}{2} \int dx\, dx' \sum_{\alpha\alpha'} \phi_{\alpha\eta x} \Pi'_{\eta\alpha\eta'\alpha'}(x - x') \phi_{\alpha'\eta' x'}, \\
\Pi'_{\eta\alpha\eta'\alpha'}(x - x') &= \langle M'_{\alpha\eta x} M'_{\alpha'\eta' x'} \rangle_{MF}.
\end{aligned} \quad (2.54)$$

With the presence of the gauge-fixing part, this action is nonsingular and unambiguously determines the bare propagator of AF $D_{\eta\alpha\eta'\alpha'}(x - x')$ by the inversion of the corresponding quadratic form. Like the parameters $v_{\mathbf{r}}$, $M_{\alpha\eta}$, the propagator $D$ is determined by this inversion on the MF level and is different from the full dressed Green's function $\langle\phi_{\alpha\eta x}\phi_{\alpha'\eta'x'}\rangle$. Since we have used RPA bubbles as a part of the free action of AF, the diagrams which include the segment shown on Fig. 1(b) must be excluded.

Obviously, the propagator $D$ is proportional to $1/N$, so the order of every diagram in the $1/N$ expansion is equal to the number of propagators $D$ minus the number of $b$- and $f$-loops and $\mathcal{L}'_g$-vertices in the diagram. Also, if we integrate out the $b$- and $f$-fields, the resulting effective action of AF will be proportional to $N$. Consequently, the order of any diagram in the $1/N$ expansion is equal to the number of AF-loops in this diagram. The MF free energy per color is of order 1, and the diagrams give to it corrections of order $(1/N)^p$, $p \geq 1$. Therefore, the MF contribution is the only one that survives the limit $N \to \infty$. As for the corrections, the simple analysis shows that there exists only a finite number of diagrams contributing to each order in $1/N$, which makes it computable at least in principle.

We now write the gauge-fixing part of Eq. (2.54) in the form

$$\int \mathcal{L}_g^{(0)}(x)\, dx = \frac{1}{2} \int dx dx' \sum_{\alpha\alpha'} \phi_{\alpha\eta x} \Pi_{\eta\alpha\eta'\alpha'}^{g\,(0)}(x - x') \phi_{\alpha'\eta'x'} , \qquad (2.55)$$

which determines the matrix $\Pi_{\eta\alpha\eta'\alpha'}^{g\,(0)}(x)$. Then in momentum representation, we have for the effective action of AF (2.54)

$$\mathcal{S}_{eff} = \frac{1}{2} \sum_q \sum_{\alpha,\alpha'} \phi_{\alpha\eta -q} \Pi_{\eta\alpha\eta'\alpha'}(q) \phi_{\alpha'\eta'q} , \qquad (2.56)$$

where the polarization matrix is

$$\begin{aligned}
\Pi_{\eta\alpha\eta'\alpha'}(q) &= \Pi_{\eta\alpha\alpha'}^{(0)} \delta_{\eta\eta'} + \Pi_{\eta\alpha\eta'\alpha'}^{g\,(0)}(q) - \Pi'_{\eta\alpha\eta'\alpha'}(q) , \\
\Pi'_{\eta\alpha\eta'\alpha'}(q) &\equiv \frac{1}{\beta V} \langle M'_{\alpha\eta q} M'_{\alpha'\eta'-q}\rangle_{MF} ,
\end{aligned} \qquad (2.57)$$

and $\Pi_{\eta\alpha\eta'\alpha'}^{g\,(0)}(q)$ is given by the Fourier transformation of $\Pi_{\eta\alpha\eta'\alpha'}^{g\,(0)}(x)$. In explicit form, the nonzero elements of $\Pi'$ in the phase with $v_{\mathbf{r}} = 0$ are

$$\Pi'_{\eta\alpha\eta'\alpha'}(q) = \begin{cases} Ne^{\frac{i}{2}(q_\eta - q_{\eta'})} \sum_k \mathrm{Tr}\left(m_{\alpha\eta\mathbf{k}} G_{k+\frac{q}{2}} m_{\alpha'\eta'\mathbf{k}} G_{k-\frac{q}{2}}\right) , & \alpha,\alpha' \in \widetilde{\mathcal{N}}_b \\ -Ne^{\frac{i}{2}(q_\eta - q_{\eta'})} \sum_k m_{\alpha\eta\mathbf{k}} g_{k+\frac{q}{2}} m_{\alpha'\eta'\mathbf{k}} g_{k-\frac{q}{2}} , & \alpha,\alpha' \in \widetilde{\mathcal{N}}_f \end{cases} \qquad (2.58)$$

where in the top (bottom) line of this formula $m_{7\mathbf{k}}$ is defined as $m^b_{7\mathbf{k}}$ ($m^f_{7\mathbf{k}}$) and $\Pi'_{77}(q)$ is equal to the sum of expressions, given by the top and bottom lines. The propagator of AF is

$$D_{\eta\alpha\eta'\alpha'}(q) = [\Pi(q)]^{-1}_{\eta\alpha\eta'\alpha'} . \qquad (2.59)$$

As was explained above, in the absence of the ordering field the same formulas (2.58) describe nonzero elements of $\Pi'$ in the phase with condensed spinons, where the substitution (2.46) has to be made if the explicit separation of condensate contribution is desired.

## III. Collective modes

The breaking of continuous symmetries implies the existence of collective modes, which we will study in this section. We will use here the formulation without the ordering field, which unifies the cases of zero and nonzero $v_{\mathbf{r}}$. The initial theory has the gauge invariance with respect to the transformations (2.19) – (2.21) with an arbitrary function $\chi_x$. We want first to consider the phase with $\mathcal{A}_\eta = 0$ and nonzero $B_\eta$, $F_\eta$. After fixing the gauge, the theory is invariant with respect to the residual symmetry, described by Eqs. (2.19) – (2.21) with $\chi_x = \chi_\eta x_\eta$. The corresponding transformation of $a_{\eta x}$ is a global shift $a_{\eta x} \to a_{\eta x} + l^{-1}\chi_\eta$, so that AF $\phi_{\alpha\eta x}$, $\alpha = 3, \ldots, 6$ undergo the global phase transformation $\phi_{\alpha\eta x} \to \exp(ig_\alpha \chi_\eta)\phi_{\alpha\eta x}$. We have therefore the global symmetry, described by a continuous group with $d$ parameters $\chi_\eta$. The nonzero values of $B_\eta$, $F_\eta$ correspond to spontaneous breaking of this symmetry. Let us show that this symmetry breaking leads to the generation of $d$ massless Goldstone modes, which are just the Cartesian components of the corresponding "photon" $a_{\eta x}$ (compare with Ref. [14]).

In the phases with a broken symmetry we denote by $\phi_{\alpha'\eta'q}$ the deviations of AF in momentum representation from their vacuum values $\beta V \phi_{\alpha'\eta'}\delta_{q0}$. We want to find the values $\phi^{(a_\eta)}_{\alpha'\eta'0}$, which correspond to the infinitesimally small field $a_{\eta x}$ (where $\eta$ is kept fixed at some particular value) with zero momentum. These values are the deviations, which result from the shift of $a_{\eta x}$ by the infinitesimally small value $\delta\chi_\eta$ from its vacuum value, if all other parameters of AF are fixed at their vacuum values. From Eq. (2.15) we have then

$$\phi^{(a_\eta)}_{\alpha'\eta'0} = \delta_{\eta'\eta} g_{\alpha'} \phi_{\alpha'\eta'}, \tag{3.1}$$

where we have dropped nonessential overall factors.

We must now prove that this mode is an eigenmode of the effective Hamiltonian of AF with zero eigenenergy. This means that $\phi^{(a_\eta)}_{\alpha'\eta'0}$ must satisfy the equation $\sum_{\alpha'} \Pi_{\eta''\alpha''\eta'\alpha'}(0)\phi^{(a_\eta)}_{\alpha'\eta'0} = 0$, with $\Pi_{\eta''\alpha''\eta'\alpha'}(q)$ given by Eq. (2.57). Since in this phase $\Pi^{g(0)}_{\eta''\alpha''\eta'\alpha'}(q)$ does not contribute for $q = 0$, we must show that

$$\sum_{\alpha'} \Pi'_{\eta''\alpha''\eta'\alpha'}(0)\phi^{(a_\eta)}_{\alpha'\eta'0} = \sum_{\alpha'} \Pi^{(0)}_{\eta''\alpha''\alpha'}\phi^{(a_\eta)}_{\alpha'\eta''0}. \tag{3.2}$$

To do this, we consider the theory with the Lagrangian

$$\mathcal{L}_K(x) \equiv \mathcal{L}_{MF}(x) - \sum_{\alpha'} K_{\alpha'\eta'x} M_{\alpha'\eta'x}, \tag{3.3}$$

where $K_{\alpha'\eta'x}$ are the sources for $M_{\alpha'\eta'x}$. Let $Z_K$ be the partition function corresponding to this Lagrangian:

$$Z_K = \int \mathcal{D}(b^*bf^*f) \exp\left[-\int \mathcal{L}_K(x)\,dx\right]. \tag{3.4}$$

We now make in the path integral in (3.4) the change of integration variables (2.19) with $\chi_x = \chi_\eta x_\eta$ (since $\eta$ is fixed, there are no summations over $\eta$ here and below!). We have then the transformation law for the composition operators $M_{\alpha'\eta'x'}$:

$$M_{\alpha'\eta'x'} \longrightarrow \exp\left\{i\chi_\eta \left[\tilde{g}_{\alpha'}\left(2x'_\eta + \delta_{\eta'\eta}\right) - g_{\alpha'}\delta_{\eta'\eta}\right]\right\} M_{\alpha'\eta'x'}. \tag{3.5}$$

We recall that $\mathcal{A}_\eta = 0$, so in $\mathcal{L}_{MF}(x)$ there are no terms with $M_{1,2\eta'x'}$.

The condition that $Z_K$ does not change with the change of integration variables yields then

$$\begin{aligned}
0 &= \left.\frac{\delta \ln Z_K}{\delta(-i\chi_\eta)}\right|_{\chi_\eta = 0} \\
&= Z_K^{-1} \int dx' \int \mathcal{D}(b^*bf^*f)\Biggl\{\sum_{\alpha'} \phi_{\alpha'\eta'} g_{\alpha'} M_{\alpha'\eta'x'}\delta_{\eta'\eta} \\
&\quad + \sum_{\alpha'} K_{\alpha'\eta'x'} M_{\alpha'\eta'x'}\left[g_{\alpha'}\delta_{\eta'\eta} - \tilde{g}_{\alpha'}(2x'_\eta + \delta_{\eta'\eta})\right]\Biggr\} \exp\left[-\int \mathcal{L}_K(x)\,dx\right].
\end{aligned} \tag{3.6}$$

The first term in the integrand will not change if we write there $M'_{\alpha'\eta'x'}$ instead of $M_{\alpha'\eta'x'}$. We then differentiate (3.6) over $K_{\alpha''\eta''x''}$ and set $K_{\alpha\eta x} = 0$ to get

$$\sum_{\alpha'} \Pi'_{\eta''\alpha''\eta'\alpha'}(0)\delta_{\eta'\eta} g_{\alpha'}\phi_{\alpha'\eta'} + \delta_{\eta''\eta} N M_{\alpha''\eta''} g_{\alpha''} = 0, \tag{3.7}$$

where we have used the self-consistensy equations and the zero value of $\mathcal{A}_\eta$ in our phase. Rewriting the second term in (3.7) as $\delta_{\eta''\eta} \sum_{\alpha'} g_{\alpha''} \Pi^{(0)}_{\eta''\alpha''\alpha'\eta'} \phi_{\alpha'\eta''} = -\delta_{\eta''\eta} \sum_{\alpha'} \Pi^{(0)}_{\eta''\alpha''\alpha'\eta'} g_{\alpha'} \phi_{\alpha'\eta''}$ we obtain the desired relation (3.2). We note the crucial importance for our proof of the absence of the terms with $M_{1,2\eta x}$ in $\mathcal{L}_{MF}$.

Consider now the phase with $B_\eta = F_\eta = 0$ and $\mathcal{A}_\eta \neq 0$. Here the residual symmetry is described by Eqs. (2.19) – (2.21) with $\chi_x = \xi_{\mathbf{r}} \chi_\eta x_\eta$. A nonzero value of $\mathcal{A}_\eta$ means that this residual symmetry is spontaneously broken; we expect that the $d$ massless Goldstone modes generated by this symmetry breaking are the components of the "staggered photon" $\tilde{a}_{\eta x}$. The form of the residual symmetry suggests that in this case the components of $\tilde{a}_{\eta x}$ with momentum $\mathbf{q} = \vec{\pi}$ have zero energy. To check this assumption we first obtain from Eq. (2.15) that the values $\phi^{(\tilde{a}_\eta)}_{\alpha'\eta'x}$ corresponding to infinitesimally small field $\tilde{a}_{\eta x}$ with momentum $\vec{\pi}$ are proportional to $\delta_{\eta'\eta} \xi_{\mathbf{r}} \tilde{g}_{\alpha'} \phi_{\alpha'\eta'}$. One can show now that the corresponding amplitudes $\phi^{(\tilde{a}_\eta)}_{\alpha'\eta'\vec{\pi}} = \delta_{\eta'\eta} \tilde{g}_{\alpha'} \phi_{\alpha'\eta'}$ satisfy the equation

$$\sum_{\alpha'} \Pi'_{\eta''\alpha''\eta'\alpha'}(0, \vec{\pi}) \phi^{(\tilde{a}_\eta)}_{\alpha'\eta'\vec{\pi}} = \sum_{\alpha'} \Pi^{(0)}_{\eta''\alpha''\alpha'} \phi^{(\tilde{a}_\eta)}_{\alpha'\eta''\vec{\pi}}, \quad (3.8)$$

which may be proven by the same method as (3.2) using the function $\chi_x = \xi_{\mathbf{r}} \chi_\eta x_\eta$ in (2.19). Since in this phase the GFT does not contribute at $\mathbf{q} = \vec{\pi}$, this means that the field $\tilde{a}_{\eta x}$ with momentum $\vec{\pi}$ has zero energy. It is again crucially important that in this phase $\mathcal{L}_{MF}$ does not have terms with $M_{\alpha\eta x}$, $\alpha = 3, \ldots, 6$.

Apriory we cannot exclude the possibility of the existence of more general "mixed" phases, which for some $\eta$ (we will denote them $\eta_u$) have $\mathcal{A}_\eta = 0$ and nonzero $B_\eta$, $F_\eta$, while for other $\eta$ (which we will denote $\eta_s$) have $\mathcal{A}_\eta \neq 0$ and $B_\eta = F_\eta = 0$. To describe these phases, it is convenient to introduce parameters $\nu_\eta$, $\tilde{\nu}_\eta$ such that $\nu_{\eta_u} = \tilde{\nu}_{\eta_s} = 1$ and $\nu_{\eta_s} = \tilde{\nu}_{\eta_u} = 0$. The corresponding Goldstone modes now are $a_{\eta_u x}$ and $\tilde{a}_{\eta_s x}$; they have zero energy in the state with momentum $\mathbf{k}^m$, where $k^m_\eta = \pi \tilde{\nu}_\eta$. To show that, one should repeat the same consideration as above, using the function $\chi_x = \exp(i \mathbf{k}^m \mathbf{r}) \chi_\eta x_\eta$. Then one obtains

$$\sum_{\alpha'} \Pi'_{\eta''\alpha''\eta'\alpha'}(0, \mathbf{k}^m) \phi^{(m\eta)}_{\alpha'\eta'\mathbf{k}^m} = \sum_{\alpha'} \Pi^{(0)}_{\eta''\alpha''\alpha'} \phi^{(m\eta)}_{\alpha'\eta''\mathbf{k}^m}, \quad (3.9)$$

with $\phi^{(m\eta)}_{\alpha'\eta'\mathbf{k}^m} = \delta_{\eta'\eta} \phi_{\alpha'\eta'} (\tilde{\nu}_{\eta'} \tilde{g}_{\alpha'} + \nu_{\eta'} g_{\alpha'})$, which proves our statement.

Now we want to consider the phase with nonzero $\mathcal{A}_\eta$ and $B_\eta$, $F_\eta$. In this phase the situation will be quite different. Indeed, the Lagrangian $\mathcal{L}_{MF}$ in this case includes the terms with $M_{\alpha\eta x}$ *for all* $\alpha = 1, \ldots, 6$. One can easily see that this results in a failure of our consideration above: Eqs. (3.2), (3.8), and (3.9) will not be satisfied. This is because in this phase the fields $\tilde{a}_{\eta x}$ and $a_{\eta x}$ got screened, obtaining a gap due to the Anderson-Higgs mechanism. Indeed, consider first the phase with $\mathcal{A}_\eta = 0$ and nonzero $B_\eta$, $F_\eta$. As we have just shown, in this phase $a_{\eta x}$ is a gapless gauge field. The state of the gauge field with zero energy corresponds to such a momentum, that the gauge transformation with this momentum does not change the gauge field. For the field $a_{\eta x}$ this momentum is equal to zero, and the corresponding function $\chi_x$ in (2.19) is a constant: $\chi_x \equiv \chi$. On the other hand, under the gauge transformation with this function $\chi_x$, the AF $\phi_{1,2\eta x}$ acquire a factor $\exp(-2i\tilde{g}_{1,2}\chi)$. Therefore, they will produce a gap in the spectrum of the gauge field $a_{\eta x}$, if we cross a border and pass into the phase with nonzero expectation value of $\phi_{1,2\eta x}$, i.e. with $\mathcal{A}_\eta \neq 0$. Similary, in the phase with $B_\eta = F_\eta = 0$ and $\mathcal{A}_\eta \neq 0$ the gapless gauge field is $\tilde{a}_{\eta x}$. Its zero energy state is the one with momentum $\vec{\pi}$, since it does not change under the gauge transfromation with a function $\chi_x = \chi \xi_{\mathbf{r}}$. Here, the fields $\phi_{\alpha\eta x}$, $\alpha = 3, \ldots, 6$ are not invariant with respect to this transformation, acquiring a factor $\exp(-2i\xi_{\mathbf{r}} g_\alpha \chi)$. Consequently, in a phase where they have nonzero expectation values, they will give a gap to the state of $\tilde{a}_{\eta x}$ with momentum $\vec{\pi}$.

When the polarization matrix $\Pi(q)$ has an eigenmode with zero eigenvalue, the inversed quantity – AF propagator $D(q)$ – has a pole. As is well known [15], the poles of the AF propagator correspond to the bound states in the original system of interacting particles, in our case – spinons and monons. In the case of $\tilde{a}_{\eta x}$ this bound state exists in the particle-particle boson channel, while in the case of $a_{\eta x}$ it exists in the particle-hole channel of both bosons and fermions. It is important to note, however, that although we found $d$ eigenmodes, only $d-1$ linear combinations of them correspond to genuine Goldstone modes. Indeed, one linear combination of these modes corresponds to the possibility of performing the gauge transformations. This mode appears in all the phases with a broken gauge symmetry; in the polarization matrix $\Pi'$ it is completely decoupled from the gauge-invariant operators and, in the absence of GFT, has zero frequency *for all momenta* [16].

To find the explicit expression for this zero mode in momentum representation we make in the path-integral (3.4) the change of integration variables (2.19) with the function $\chi_x = \chi \exp(iq_\eta x_\eta)$, where this time the summation over $\eta$ is assumed. For small $\chi$, the variations of different terms of $\mathcal{L}_K$ are given by

$$\delta M_{\alpha'\eta'x'} = \chi l e^{iq_\eta x_\eta} \left( \tilde{d}_{q\eta'} \tilde{g}_{\alpha'} + d_{q\eta'} g_{\alpha'} \right) M_{\alpha'\eta'x'}, \tag{3.10}$$

where $d_{q\eta}$ and $\tilde{d}_{q\eta}$ are the Fourier images of the usual and staggered lattice derivatives:

$$\begin{aligned}
d_{q\eta} &\equiv \frac{i}{l}\left(1 - e^{iq_\eta}\right) = \frac{2}{l} e^{i\frac{q_\eta}{2}} \sin\frac{q_\eta}{2} \xrightarrow[q_\eta \to 0]{} \frac{q_\eta}{l}, \\
\tilde{d}_{q\eta} &\equiv \frac{i}{l}\left(1 + e^{iq_\eta}\right) = \frac{2i}{l} e^{i\frac{q_\eta}{2}} \cos\frac{q_\eta}{2}.
\end{aligned} \tag{3.11}$$

Therefore, we obtain by the same method

$$\sum_{\alpha'} \left[ \Pi^{(0)}_{\eta''\alpha''\alpha'} \delta_{\eta''\eta'} - \Pi'_{\eta''\alpha''\eta'\alpha'}(0, \mathbf{q}) \right] \phi^{(0)}_{\alpha'\eta'\mathbf{q}} = 0, \tag{3.12}$$

$$\phi^{(0)}_{\alpha'\eta'\mathbf{q}} = \left( \tilde{d}_{q\eta'} \tilde{g}_{\alpha'} + d_{q\eta'} g_{\alpha'} \right) \phi_{\alpha'\eta'}. \tag{3.13}$$

These equations show that for every $\mathbf{q}$ in the absence of GFT the mode $\phi^{(0)}_{\alpha\eta\mathbf{q}}$ is an eigenmode with zero eigenfrequency. A different derivation of Eq. (3.12), using the properties of the $b$ and $f$ Green's functions, will be presented in Appendix.

Equations (3.12) – (3.13) can be also used for the alternative derivation of the results, discussed above. Indeed, if for every $\eta$ only $\phi_{1,2\eta}$ or $\phi_{\alpha\eta}$, $\alpha = 3,\ldots,6$ are different from zero, then we can obtain (3.9) from (3.12) by differentiating (3.12) over $q_\eta$ and setting $\mathbf{q} = \mathbf{k}^m$. If, on the other hand, at least for one $\eta$ both $\phi_{1,2\eta}$ and $\phi_{\alpha\eta}$, $\alpha = 3,\ldots,6$ are nonzero, then this derivation does not work, and the only mode with zero frequency is the zero mode $\phi^{(0)}_{\alpha\eta\mathbf{q}}$.

Now let $\mathcal{R}$ be an arbitrary gauge-invariant operator. By adding the term $K\mathcal{R}$ to the Lagrangian $\mathcal{L}_{MF}$ and considering the variation of $Z_K$ (3.4) with respect to $\chi$ and $K$, it is easy to get

$$\langle \mathcal{R} \sum_\alpha \phi^{(0)}_{\alpha\eta\mathbf{q}} M'_{\alpha\eta\mathbf{q}} \rangle_{MF} = 0, \tag{3.14}$$

thus proving that the zero mode and the gauge-invariant quantities are not coupled through the matrix $\Pi'$.

Now we want to make some comments on our choice of the GFT. In the real space, the states of the zero mode have the amplitudes $\phi_{\alpha'\eta'x'}$, which are equal to the deviations of AF from their vacuum values $\phi_{\alpha'\eta'}$, generated by the gauge transformations (2.19) – (2.21) with arbitrary functions $\chi_{x'}$. For nonzero $\epsilon$, these functions must be time-independent. A convenient basis set of them is given by the $\delta$-functions $\chi^{(\mathbf{r})}_{x'} \equiv \chi \delta_{\mathbf{r}'\mathbf{r}}$. From Eqs. (2.15) and (2.19) – (2.21) we have for small $\chi$ the expression for the corresponding zero-mode amplitudes $\phi^{(\mathbf{r})}_{\alpha'\eta'x'}$:

$$\phi^{(\mathbf{r})}_{\alpha'\eta'x'} = il\chi \phi_{\alpha'\eta'} \left( \tilde{g}_{\alpha'} \tilde{\partial}^R_{\eta'} + g_{\alpha'} \partial^R_{\eta'} \right) \delta_{\mathbf{r}'\mathbf{r}}. \tag{3.15}$$

In the space of AF, taken at some particular moment of imaginary time $\tau$, the scalar product is defined as

$$\begin{aligned}
\left( \phi^{(1)}, \phi^{(2)} \right)_\tau &\equiv \sum_{\mathbf{r}'} \left[ \frac{1}{2} \sum_{\alpha=1}^{6} \left( \phi^{(1)'}_{\alpha\eta x'} \phi^{(2)'}_{\alpha\eta x'} + \phi^{(1)''}_{\alpha\eta x'} \phi^{(2)''}_{\alpha\eta x'} \right) + \phi^{(1)}_{7x'} \phi^{(2)}_{7x'} \right] \\
&= \frac{1}{2} \sum_{\alpha\beta} \sum_{\mathbf{r}'} \phi^{(1)}_{\alpha\eta x'} \gamma_{\alpha\beta} \phi^{(2)}_{\beta\eta x'},
\end{aligned} \tag{3.16}$$

where in the right hand side of this equation $x' = (\tau, \mathbf{r}')$, $\phi^{(1,2)'}_{\alpha\eta x'}$ and $\phi^{(1,2)''}_{\alpha\eta x'}$ are the real and imaginary parts of the AF $\phi^{(1,2)}_{\alpha\eta x'}$, $\gamma_{\alpha\beta}$ is a matrix with nonzero elements $\gamma_{12} = \gamma_{21} = \gamma_{35} = \gamma_{53} = \gamma_{46} = \gamma_{64} = \gamma_{77}/2 = 1$, and we have used Eq. (2.12). In the absence of the GFT, the quadratic form of the effective free Lagrangian of AF in Eq. (2.54) does not depend on the components of AF along the zero mode states $\phi^{(\mathbf{r})}$. These components are equal to the scalar product $(\phi, \phi^{(\mathbf{r})})_\tau$. The lattice derivatives $\partial_\eta^{R,L}$, $\tilde{\partial}_\eta^{R,L}$ allow integration by parts:

$$\sum_{\mathbf{r}} f_x \, \nabla_\eta^R \, g_x = -\sum_{\mathbf{r}} g_x \, \nabla_\eta^L \, f_x \,, \tag{3.17}$$

where $\nabla_\eta$ stands for both $\partial_\eta$ and $\tilde{\partial}_\eta$ and the functions $f_x$, $g_x$ satisfy the periodic boundary conditions (or the lattice is assumed to be infinite). Now using the expression (3.15) for $\phi^{(\mathbf{r})}_{\alpha'\eta'x'}$ and performing the integration by parts in the definition (3.16) of the scalar product we obtain the equality

$$L_x^{(0)} = -\frac{1}{2\, l^2 \chi} \left( \phi, \phi^{(\mathbf{r})} \right)_\tau . \tag{3.18}$$

Therefore, the quadratic part (2.53) of the GFT gives the potential energy to the component of AF, which originally (i.e., without gauge fixing) did not have potential energy because of the gauge invariance. We note that one can also regard $\mathcal{L}_g^{(0)}(x)$, Eq. (2.53), as the GFT, with $\mathcal{L}_g'(x)$ being the logarithm of the corresponding Faddeev-Popov determinant. In the limit $\zeta \to 0$ this leads us to the formulation used in [16, 17]. Keeping $\zeta$ finite, however, allows us to obtain explicit expressions for all contributions, corresponding to the gauge-fixing procedure, which will be important for us in the forthcoming sections. We also note here, that rather then refer to the finite $\epsilon$ in order to prohibit the time-dependent gauge transformations, one can eliminate the residual gauge freedom with respect to such transformations by adding to $L_x$ (2.49) the term $(2l^2)^{-1}\partial_0 a_{0x}$. Then for the corresponding new $L_x^{(0)}$, Eq. (3.18) will remain correct, if in its right hand side $\left(\phi, \phi^{(\mathbf{r})}\right)_\tau$ is changed to $\left(\phi, \phi^{(x)}\right)$, where

$$\phi^{(x)}_{\alpha'\eta'x'} = \chi \left[ il\phi_{\alpha'\eta'} \left( \tilde{g}_{\alpha'}\tilde{\partial}_{\eta'}^R + g_{\alpha'}\partial_{\eta'}^R \right) + \delta_{\alpha'7}\frac{\partial}{\partial\tau'} \right] \delta_{\mathbf{r}'\mathbf{r}}\delta(\tau'-\tau) \tag{3.19}$$

is a zero-mode amplitude, generated by the time-dependent $\delta$-function $\chi_{x'}^{(x)} \equiv \chi\delta_{\mathbf{r}'\mathbf{r}}\delta(\tau'-\tau)$ and

$$\left( \phi^{(1)}, \phi^{(2)} \right) \equiv \frac{1}{2}\sum_{\alpha\beta} \int dx' \, \phi^{(1)}_{\alpha\eta x'} \gamma_{\alpha\beta} \phi^{(2)}_{\beta\eta x'} \tag{3.20}$$

is a scalar product of AF, taken on the whole interval $(0, \beta)$ of the variation of imaginary time.

## IV. Equations of motion

The variation of the Lagrangian (2.13) with respect to AF yields the useful equations of motion

$$M'_{\alpha\eta x} = \sum_\beta \Pi^{(0)}_{\eta\alpha\beta}\phi_{\beta\eta x}. \tag{4.1}$$

The use of these equations, however, requires special care and in a simple form (4.1) is possible only under special conditions. Indeed, our diagram technique is produced by the Lagrangian (2.22), rather than (2.13). To elucidate the possibility of using Eq. (4.1) in our diagram technique, we start with the obvious equality

$$\int \mathcal{D}(b^*b f^* f \phi) \frac{\delta}{\delta \phi_{\alpha\eta x}} e^{-\int \mathcal{L}(x)\,dx} = 0. \tag{4.2}$$

With the Lagrangian (2.22), Eq. (4.2) leads to the identities

$$\langle M'_{\alpha\eta x} \rangle = \Big\langle \sum_\beta \Pi^{(0)}_{\eta\alpha\beta}\phi_{\beta\eta x} + \frac{\delta \mathcal{S}_g}{\delta \phi_{\alpha\eta x}} \Big\rangle, \tag{4.3}$$

which are just the manifestation of the equations of motion in our model. In the last term in (4.3) $\mathcal{S}_g \equiv \int \mathcal{L}_g(x)dx$ is the gauge-fixing part of the action. Here we want to show how our diagram technique meets (4.3). We will employ the generalization of the method, used in [9]. We start with the observation that the operator $M'_{\alpha\eta x}$ in the left hand side of (4.3) can produce three kinds of (sub)diagrams, presented in Fig. 2(a–c). On the diagram of Fig. 2(a) the $b$-$f$ lines, leaving $M'_{\alpha\eta x}$, go to different vertices. On the diagram of Fig. 2(b) these lines, after leaving $M'_{\alpha\eta x}$, meet again in the vertex of interaction with AF, while the AF then propagates till the next vertex of interaction with the $b$-$f$ line. Since this last vertex is produced by the action of $\mathcal{L}_{int}$, and not by an application of the external field, the $b$-$f$ lines, leaving this vertex, must then go to different vertices. Physically, the first diagram describes the bare action of $M'_{\alpha\eta x}$, while the second – its renormalization due to the screening. Finally, on the diagram of Fig. 2(c) AF propagates till the vertex produced by $\mathcal{L}'_g(x)$. The analytic expression corresponding to this vertex is $-\partial \mathcal{S}'_g/\partial \phi_{\alpha''\eta''x''}$, where $\mathcal{S}'_g \equiv \int \mathcal{L}'_g(x)dx$ and $\alpha''\eta''x''$ is the label of the AF line entering this vertex.

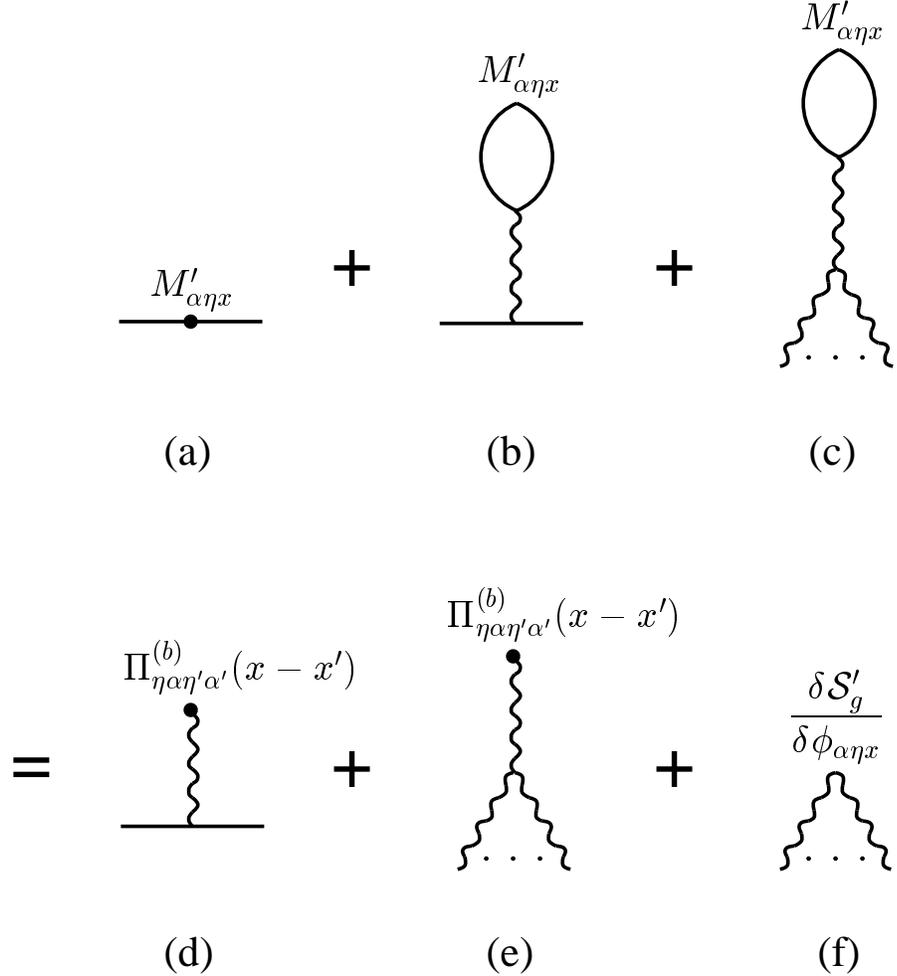

FIG. 2. Diagrammatic representation of the equations of motion, Eq. (4.3).

Now introducing the "bare" polarization matrix

$$\Pi^{(b)}_{\eta\alpha\eta'\alpha'}(x-x') \equiv \Pi^{(0)}_{\eta\alpha\alpha'}\delta_{\eta\eta'}\delta(x-x') + \Pi^{g\,(0)}_{\eta\alpha\eta'\alpha'}(x-x') \tag{4.4}$$

and using the relation

$$\Pi'_{\eta\alpha\eta'\alpha'}(x-x') = \Pi^{(b)}_{\eta\alpha\eta'\alpha'}(x-x') - \left(D^{-1}\right)_{\eta\alpha\eta'\alpha'}(x-x') \tag{4.5}$$

between the polarization matrices and the AF propagator, we can see that the sum of the diagrams Fig. 2(a–c) is equal to the sum of the diagrams Fig. 2(d–f). Since this latter sum represents the action of the right hand side of Eq. (4.3), we conclude that the calculations in our diagram technique automatically satisfy the equations of motion (4.3).

By the same logic one can prove the more general identity

$$\langle \mathcal{F} M'_{\alpha\eta x} \rangle = \Big\langle \mathcal{F} \sum_\beta \Pi^{(0)}_{\eta\alpha\beta} \phi_{\beta\eta x} + \mathcal{F} \frac{\delta \mathcal{S}_g}{\delta \phi_{\alpha\eta x}} \Big\rangle - \Big\langle \frac{\delta \mathcal{F}}{\delta \phi_{\alpha\eta x}} \Big\rangle, \qquad (4.6)$$

where $\mathcal{F}$ is an arbitrary functional of AF and the fields $b^*$, $b$, $f^*$, $f$. In the diagrammatic proof of Eq. (4.6) the diagram of Fig. 3, which exists when $\mathcal{F}$ depends on AF, should be added to the diagrams of Fig. 2(a–c). It is important that since all the diagrams of Fig. 2 are of the same order in $1/N$, the identities (4.3), (4.6) are satisfied in each order of the $1/N$ expansion separately. In the special case of $\alpha = 7$ and $\mathcal{F}$ independent of $\phi_{7x}$, (4.6) gives $N^{-1} \langle \mathcal{F} M_{7x} \rangle = \mathcal{O}(\epsilon)$; the factor $N^{-1}$ here cancels the summation over colors in $M_{7x}$. We conclude that in the limit $\epsilon \to 0$ our diagram technique satisfies the constraint (2.10) as an operator identity in each order of the $1/N$ expansion separately.

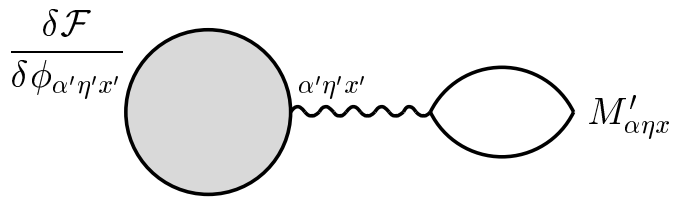

FIG. 3. The diagram for $\langle \mathcal{F} M'_{\alpha\eta x} \rangle$, which should be added to the diagrams of Fig. 2(a–c) in the case when $\mathcal{F}$ depends on AF. The shaded circle represents the diagrams for $\delta \mathcal{F}/\delta \phi_{\alpha'\eta'x'}$, where $\alpha'\eta'x'$ is the label of the AF line entering the circle.

Now we want to show that in many important cases the term with $\delta \mathcal{S}_g/\delta \phi_{\alpha\eta x}$ in (4.6) gives vanishing contribution. We first show that $L_x$, Eq. (2.49), is decoupled from the gauge-invariant quantities:

$$\langle L_x \mathcal{R} \rangle = 0 \qquad (4.7)$$

for arbitrary gauge-invariant functional $\mathcal{R}$. To prove this identity, we add to the action the term $K\mathcal{R}$, where $K$ is a source of $\mathcal{R}$, and make in the path-integral representation for the partition function the change of integration variables, corresponding to the gauge transformation with the infinitesimally small parameter $\chi_x$. In the state with a broken symmetry this change is given by the formulas

$$\begin{aligned}
b_{\sigma x} &\longrightarrow b_{\sigma x} + i\chi_x (v_{\sigma \mathbf{r}} + b_{\sigma x}), \\
\phi_{\alpha\eta x} &\longrightarrow \phi_{\alpha\eta x} + il \left( \phi_{\alpha\eta} + \phi_{\alpha\eta x} \right) \left( \tilde{g}_\alpha \tilde{\partial}^R_\eta + g_\alpha \partial^R_\eta \right) \chi_x, \quad \alpha = 1, \ldots, 6
\end{aligned} \qquad (4.8)$$

and the old formulas (2.19) – (2.20) for the transformation of $f_x$ and $\phi_{7x}$. The variation of the gauge-fixing part of the action under this transformation is

$$\begin{aligned}
\delta \mathcal{S}_g &= \frac{N}{\zeta} \int dx \, L_x \tilde{\triangle} \chi_x, \\
\tilde{\triangle} &\equiv \frac{1}{8} J^2_\eta |\mathcal{A}_\eta|^2 \tilde{\partial}^L_\eta \tilde{\partial}^R_\eta + t^2_\eta B_\eta F_\eta \partial^L_\eta \partial^R_\eta.
\end{aligned} \qquad (4.9)$$

Since the rest of the action is gauge-invariant, the condition that partition function does not change with a change of integration variables yields

$$0 = \delta Z = -\int \mathcal{D}(b^* b f^* f \phi) \delta \mathcal{S}_g \exp(-\mathcal{S} + K\mathcal{R}), \qquad (4.10)$$

where $\delta \mathcal{S}_g$ is given by (4.9). Now differentiating this equality over $\tilde{\triangle}\chi_x$ and $K$, and setting $K=0$, we obtain (4.7). Along with $\langle L_x \mathcal{R}\rangle$, the derivatives of $\langle L_x \mathcal{R}\rangle$ with respect to $x$ also vanish.

Now returning to Eq. (4.6), we have there after integration by parts and using Eq. (2.51)

$$\frac{\delta \mathcal{S}_g}{\delta \phi_{\alpha \eta x}} = -\frac{N}{16il\zeta} \frac{J_\eta^2 |\mathcal{A}_\eta|^2 \tilde{g}_\alpha \tilde{\partial}_\eta^R + 4t_\eta^2 B_\eta F_\eta g_\alpha \partial_\eta^R}{\phi_{\alpha \eta} + \phi_{\alpha \eta x}} L_x \; , \qquad (4.11)$$

where the term $(\phi_{\alpha \eta} + \phi_{\alpha \eta x})^{-1}$ is understood as a power series. Under the gauge transformation (4.8), the term $\phi_{\alpha \eta} + \phi_{\alpha \eta x}$ in (4.11) obtains the factor $\exp[il(\tilde{g}_\alpha \tilde{\partial}_\eta^R + g_\alpha \partial_\eta^R)\chi_x]$. Therefore, if the functional $\mathcal{F}$ in (4.6) obtains the same factor, then as was just shown, the term $\mathcal{F}\delta \mathcal{S}_g/\delta \phi_{\alpha \eta x}$ there may be neglected. For example, it will be so in the important case when $\mathcal{F} = M_{\beta \eta x}$ with $\beta$ such that $h_{\eta \alpha \beta} \neq 0$. In such cases, instead of Eq. (4.6) we can use a simpler relation

$$\langle \mathcal{F} M'_{\alpha \eta x}\rangle = \Big\langle \mathcal{F} \sum_\beta \Pi^{(0)}_{\eta \alpha \beta} \phi_{\beta \eta x}\Big\rangle - \Big\langle \frac{\delta \mathcal{F}}{\delta \phi_{\alpha \eta x}}\Big\rangle. \qquad (4.12)$$

This simplification, allowing us to neglect the contribution of $\delta \mathcal{S}_g/\delta \phi_{\alpha \eta x}$ in Eq. (4.6), is a usefull property of our GFT. Below, when we use Eq. (4.12), it is always assumed that the conditions for neglecting the term with $\delta \mathcal{S}_g/\delta \phi_{\alpha \eta x}$ in (4.6) are satisfied, so we will omit the corresponding discussion. When $\delta \mathcal{F}/\delta \phi_{\alpha \eta x} = 0$, Eq. (4.12) allows us to use the equations of motion (4.1), following from the Lagrangian in (2.13), as operator identities. The meaning of this rule is that the total (screened) action of $M'_{\alpha \eta x}$ comes from the creation of the intermediate state of the $b$-$f$ plasmon. The plasmon is represented by the set of AF, and the amplitude of transformation of $M'_{\alpha \eta x}$ to $\phi_{\beta \eta x}$ is equal to $\Pi^{(0)}_{\eta \alpha \beta}$.

## V. Interaction with electromagnetic field

We now want to introduce to the model the interaction with the external classical electromagnetic field. The interaction with the scalar potential $\varphi(\mathbf{r})$ is described by adding to the Hamiltonian (2.1) the term

$$e\sum_\mathbf{r} \varphi(\mathbf{r}) \psi^\dagger_{\sigma \mathbf{r}} \psi_{\sigma \mathbf{r}} = \frac{e}{2} \sum_\mathbf{r} \varphi(\mathbf{r}) \left( b^\dagger_\mathbf{r} b_\mathbf{r} + f^\dagger_\mathbf{r} f_\mathbf{r} \right) \; , \qquad (5.1)$$

where $e$ is the electron's charge and we have used the constraint to make the transformation. The generalization $1 \to N$ does not affect this form of interaction. The interaction with the vector potential $\mathbf{A}(\mathbf{r})$ is obtained in a standard way [5] by modifying the hopping term $H_t$ in (2.5) as

$$H_t \longrightarrow -\frac{t_\eta}{N} \sum_\mathbf{r} \left[ e^{ielA_\eta(\mathbf{r})} b^\dagger_{\sigma \mathbf{r}+\hat{\eta}} b_{\sigma \mathbf{r}} f^\dagger_{\mathbf{r}+\hat{\eta}} f_\mathbf{r} + e^{-ielA_\eta(\mathbf{r})} b^\dagger_{\sigma \mathbf{r}} b_{\sigma \mathbf{r}+\hat{\eta}} f^\dagger_\mathbf{r} f_{\mathbf{r}+\hat{\eta}} \right] \; , \qquad (5.2)$$

where we made the $1 \to N$ generalization. The electric current is defined by the usual relation $j^e_\eta(\mathbf{r}) = -\delta H/\delta A_\eta(\mathbf{r})$. Up to terms of first order in $\mathbf{A}$, we have $\mathbf{j}^e(\mathbf{r}) = e\mathbf{j}(\mathbf{r}) + \mathbf{j}^d(\mathbf{r})$, where $\mathbf{j}(\mathbf{r})$ is the electron current, while $e\mathbf{j}(\mathbf{r}) = \mathbf{j}^p(\mathbf{r})$ and $\mathbf{j}^d(\mathbf{r})$ are respectively the paramagnetic and diamagnetic parts of the electric current. We have

$$j_\eta(\mathbf{r}) = \frac{ilt_\eta}{N} \left( b^\dagger_{\sigma \mathbf{r}+\hat{\eta}} b_{\sigma \mathbf{r}} f^\dagger_{\mathbf{r}+\hat{\eta}} f_\mathbf{r} - b^\dagger_{\sigma \mathbf{r}} b_{\sigma \mathbf{r}+\hat{\eta}} f^\dagger_\mathbf{r} f_{\mathbf{r}+\hat{\eta}} \right) \; , \qquad (5.3)$$

$$j^d_\eta(\mathbf{r}) = -\frac{e^2 l^2 t_\eta}{N} A_\eta(\mathbf{r}) \left( b^\dagger_{\sigma \mathbf{r}+\hat{\eta}} b_{\sigma \mathbf{r}} f^\dagger_{\mathbf{r}+\hat{\eta}} f_\mathbf{r} + b^\dagger_{\sigma \mathbf{r}} b_{\sigma \mathbf{r}+\hat{\eta}} f^\dagger_\mathbf{r} f_{\mathbf{r}+\hat{\eta}} \right) \; . \qquad (5.4)$$

In terms of the operators $M_{\alpha \eta \mathbf{r}}$, the electron current is

$$j_\eta(\mathbf{r}) = \frac{il}{2N} \sum_{\alpha \beta} M_{\alpha \eta \mathbf{r}} j_\alpha h_{\eta \alpha \beta} M_{\beta \eta \mathbf{r}} \; , \qquad (5.5)$$

where the $j_\alpha$ were determined in Eq. (2.16). Due to the presence of the EMF the action obtains an additional contribution

$$\mathcal{S}_{em} = \int dx \left\{ \rho(x)\varphi(x) - \left[ j_\eta^p(x) + \frac{1}{2} j_\eta^d(x) \right] A_\eta(x) \right\}, \tag{5.6}$$

where we have neglected the terms of third and higher powers in $elA_\eta$, and have denoted $\rho(x) \equiv en_x$.

It is convenient to make in (5.5) the substitution (4.1) and obtain

$$\begin{aligned} j_\eta(x) &= \frac{il}{2} \sum_\alpha j_\alpha M_{\alpha\eta x}(\phi_{\alpha\eta} + \phi_{\alpha\eta x}) \\ &\equiv \frac{1}{2} \left[ J_\eta^b(x) + J_\eta^f(x) \right], \end{aligned} \tag{5.7}$$

where we have also used (2.24) and have defined

$$J_\eta^{b,f}(x) = il \sum_{\alpha_{b,f}} j_\alpha M_{\alpha\eta x}(\phi_{\alpha\eta} + \phi_{\alpha\eta x}). \tag{5.8}$$

The currents $J_\eta^{b,f}(x)$ defined in (5.8) are invariant with respect to the gauge transformation (2.19) – (2.21); one can identify them with the currents of spinons and monons. We will soon see that this identification is not useful, but first we will consider these currents in more detail. We note that these currents may be also obtained in a different way. Namely, we can modify the Lagrangian in (2.11) in the same way as in (5.2), and then decouple it using AF $\phi_{\alpha\eta x}$. The factors $\exp(\pm ielA_{\eta x})$ should then be decoupled to the products of $\exp(\pm ie\kappa_b lA_{\eta x})$ and $\exp(\pm ie\kappa_f lA_{\eta x})$, where $\kappa_b$ ($\kappa_f$) is a part of the electron charge, belonging to the spinon (monon). All results must be independent of the choice of $\kappa_{b,f}$, if they satisfy the condition $\kappa_b + \kappa_f = 1$. The term $\sum_\alpha \phi_{\alpha\eta x} M_{\alpha\eta x}$ in (2.13) will then obtain the form $\sum_\alpha \exp(ielj_\alpha \kappa_\alpha A_{\eta x})\phi_{\alpha\eta x} M_{\alpha\eta x}$, where $\kappa_\alpha = \kappa_b(\kappa_f)$ for $\alpha \in \mathcal{N}_b(\mathcal{N}_f)$. Now differentiating this Lagrangian with respect to $-eA_{\eta x}$ for $\mathbf{A} = 0$, and performing the shift of AF, we obtain for the electron current

$$j_\eta(x) = \kappa_b J_\eta^b(x) + \kappa_f J_\eta^f(x). \tag{5.9}$$

Equation (5.7) corresponds to the symmetric choice $\kappa_b = \kappa_f = 1/2$.

For the current (5.9) to be independent of the choice of $\kappa_{b,f}$, we must have

$$J_\eta^b(x) = J_\eta^f(x) \tag{5.10}$$

as an operator identity. It is easy to see that this equality is a consequence of the absence of nontrivial bare kinetic energy of AF in our theory. Indeed, we can for example consider the variation of the Lagrangian (2.13) with respect to the transformation $a_{\eta x} \to a_{\eta x} + \delta a_{\eta x}$ and use the equation of motion $\delta \mathcal{L}/\delta a_{\eta x} = 0$. Under this transformation, the variations of AF in (2.13) are $\delta \phi_{\alpha\eta x} = ilg_\alpha \phi_{\alpha\eta x} \delta a_{\eta x}$. Performing now the differentiation of the Lagrangian (2.13) over $a_{\eta x}$ and a shift of AF, we obtain

$$0 = \delta \mathcal{L}/\delta a_{\eta x} = -il \sum_\alpha g_\alpha (\phi_{\alpha\eta} + \phi_{\alpha\eta x}) M_{\alpha\eta x}. \tag{5.11}$$

The last expression here is equal to $J_\eta^f(x) - J_\eta^b(x)$, so we immediately obtain (5.10). This also implies $j_\eta(x) = J_\eta^{b,f}(x)$, in agreement with [18]. If AF had a nontrivial bare kinetic energy, then there would have been an additional term on the right hand side of (5.11), which would violate the equality (5.10). On the other hand, we can show that (5.10) is a consequence of the equations of motion, considered in the previous section. Indeed, using (4.12) on the right hand side of (5.11), we get

$$J_\eta^f(x) - J_\eta^b(x) = -il \sum_{\alpha\beta}(\phi_{\alpha\eta} + \phi_{\alpha\eta x})g_\alpha \Pi_{\eta\alpha\beta}^{(0)}(\phi_{\beta\eta} + \phi_{\beta\eta x}) + il \sum_\alpha g_\alpha, \tag{5.12}$$

which vanishes due to (2.16) and the antisymmetry of $g_\alpha \Pi_{\eta\alpha\beta}^{(0)}$. We see that although (5.10) looks like an *equation*, which puts nontrivial restrictions on the observable quantities in our theory, it is in fact an *identity*, which is satisfied *automatically* due to the equations of motion in every order of the $1/N$ expansion and does not provide any restrictions. In other words, $J_\eta^{b,f}(x)$ are just the different forms of the electron current, and we will not associate them with the currents of spinons and monons.

A convenient representation of the electron current may be obtained by the separation of different terms in (5.7). We substitute there $M_{\alpha\eta x} = NM_{\alpha\eta} + M'_{\alpha\eta x}$ and $\phi_{\alpha\eta} = \sum_\beta h_{\eta\alpha\beta} M_{\beta\eta}$ and use (2.40) to get

$$j_\eta(x) = \frac{1}{2}\left[j_\eta^b(x) + j_\eta^f(x) + j_\eta^\phi(x) + j_\eta^s(x)\right], \tag{5.13}$$

$$j_\eta^{b,f}(x) = ilt_\eta M_{f,b\eta} \sum_{\alpha_{b,f}} j_\alpha M'_{\alpha\eta x}, \tag{5.14}$$

$$j_\eta^\phi(x) = ilN \sum_\alpha j_\alpha M_{\alpha\eta}\phi_{\alpha\eta x}, \tag{5.15}$$

$$j_\eta^s(x) = il \sum_\alpha j_\alpha M'_{\alpha\eta x}\phi_{\alpha\eta x}. \tag{5.16}$$

The formulas (5.14) may be written in the familiar form

$$\begin{aligned} j_\eta^b(x) &= \frac{i}{2m_\eta^b}\left(\partial_\eta^R b_{\sigma x}^* \cdot b_{\sigma x} - b_{\sigma x}^* \cdot \partial_\eta^R b_{\sigma x}\right) \\ j_\eta^f(x) &= \frac{i}{2m_\eta^f}\left(\partial_\eta^R f_x^* \cdot f_x - f_x^* \cdot \partial_\eta^R f_x\right) \end{aligned} \tag{5.17}$$

where the substitution $b_{\sigma x} \to v_{\sigma \mathbf{r}} + b_{\sigma x}$ should be made in the equation for $j_\eta^b(x)$ if $v_\mathbf{r} \neq 0$, and where we have denoted

$$m_\eta^{b,f} = \frac{1}{2l^2 t_\eta M_{f,b\eta}}. \tag{5.18}$$

Therefore, the hopping matrix element of the monon (spinon) determines the effective mass of the spinon (monon); zero hopping corresponds to an infinite effective mass.

The factor $1/2$ in (5.13) corresponds to the symmetric choice $\kappa_b = \kappa_f = 1/2$. If we had made a different choice, we wouldn't have this factor in (5.13), but instead every term in the sums over $\alpha$ in (5.14) – (5.16) would have obtained a factor $\kappa_\alpha$. As we discussed above, these changes would not affect any physical results.

By the equations of motion (4.3), the current $j_\eta^\phi(x)$ is different from $j_\eta^b(x) + j_\eta^f(x)$ only by the contribution of the GFT. As is clear from Eqs. (4.11) and (4.7) this contribution is always of higher order in $1/N$ and may be neglected on the MF level, $N \to \infty$. Also of higher order is the contribution of $j_\eta^s(x)$. We conclude that on the MF level the expression for the total electron current is

$$j_\eta^{MF}(x) = j_\eta^b(x) + j_\eta^f(x), \tag{5.19}$$

which corresponds to the independent interaction of EMF with the currents of spinons and monons (5.17). One can easily convince himself that representation (5.19) remains correct for every choice of $\kappa_{b,f}$.

Now let us consider the interaction with EMF, which is produced by the density $\rho(x)$ and the currents $j_\eta^{b,f,\phi}(x)$, Eqs. (5.14) – (5.15). We define the corresponding effective vertex of interaction of the $b$-$f$ line with EMF, represented by the diagram of Fig. 4(a), in such a way that the disconnection of EMF from this vertex would leave us with a legitimate diagram. Two obvious terms of the effective vertex correspond to the diagrams of Fig. 4(b) (representing the "direct" contribution of $\rho(x)$ and $j^{b,f}(x)$) and Fig. 4(c) (representing the contribution of $j_\eta^\phi(x)$). However, apart from these terms, the effective vertex must also include the "screening" term, represented by the diagram of Fig. 4(d), which describes the screened contribution of $\rho(x)$ and $j^{b,f}(x)$. Indeed, the disconnection of EMF from the diagram of Fig. 4(d) would leave us with the forbidden subdiagram of Fig. 1(a). We note, that although the diagram of Fig. 1(b) is forbidden, the diagrams of Fig. 5(a,b) are not, and should be taken into account. The diagrams of Fig. 5(c,d) will be discussed later.

Because of the presence of the GFT, we must also introduce the effective vertex of interaction of EMF with AF. This vertex is presented in the Fig. 6. We will denote this vertex as $V_\mu^{(n)}(q)$, where its dependence on momenta other than the momentum $q$ of the EMF is made implicit and the number $n$ of AF, leaving this vertex on the left hand side diagram of Fig. 6, must be greater than or equal to 2. The $n + 1 -$ particle interaction vertex on the right hand side diagrams of Fig. 6 is produced by the interaction part of the GFT, $\mathcal{L}'_g(x)$.

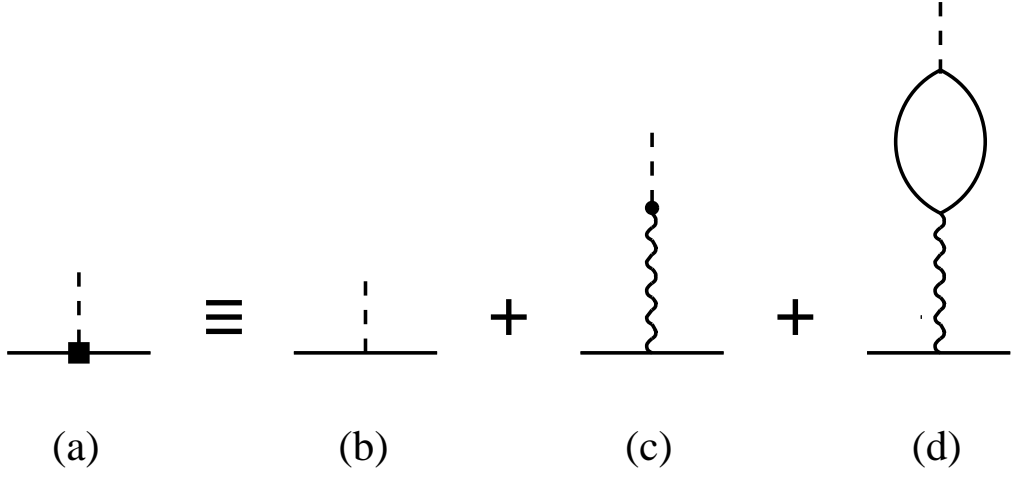

FIG. 4. The effective vertex of interaction of the $b$-$f$ line with the EMF, corresponding to the currents $j_\eta^{b,f,\phi}(x)$, Eqs. (5.14 − 5.15). The dashed tail represents the EMF.

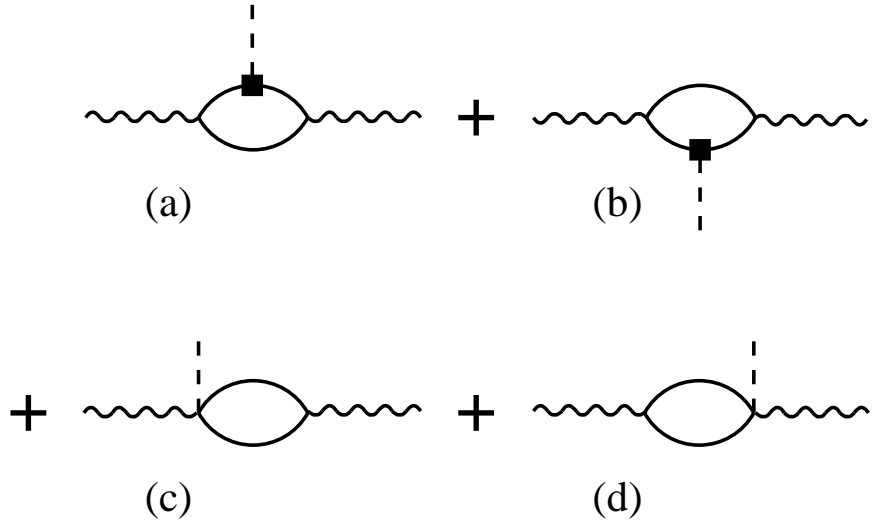

FIG. 5. The (sub)diagrams which become forbidden ones after the disconnection of the EMF line.

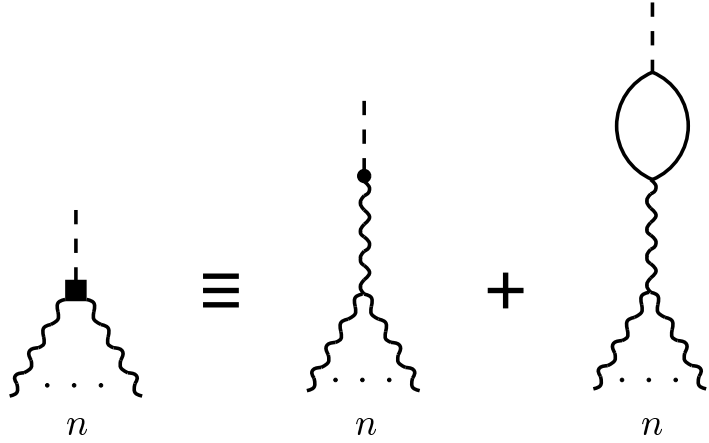

FIG. 6. The effective vertex of interaction of the AF with the EMF; $n$ is the number of the AF lines leaving the vertex.

The analytic expression corresponding to the segment of Fig. 4(a) is

$$eF^{b,f}_{k-\frac{q}{2}} \left[ \varphi(-q) V^{b,f}_0(q,k) - A_\eta(-q) V^{b,f}_\eta(q,k) \right] F^{b,f}_{k+\frac{q}{2}}, \tag{5.20}$$

where $F^{b,f}$ are the corresponding Green's functions: $F^b_p \equiv G_p$ and $F^f_p \equiv g_p$. The interaction vertices here are sums of the contributions of the diagrams of Fig. 4(b – d), $V^{b,f}_\mu(q,k) = \sum_i V^{b,f(i)}_\mu(q,k)$, where the index $i = b, c, d$ corresponds to the diagrams of Fig. 4. By Eqs. (5.6), (5.13) – (5.15), and (2.31) we have

$$\begin{aligned}
V^{b,f(b)}_0(q,k) &= \frac{1}{2} \\
V^{b,f(b)}_\eta(q,k) &= \frac{il}{2} t_\eta M_{f,b\eta} e^{i\frac{q\eta}{2}} \sum_{\alpha_{b,f}} j_\alpha m_{\alpha\eta \mathbf{k}} = \frac{il}{2} e^{i\frac{q\eta}{2}} \sum_{\alpha_{b,f}} j_\alpha \phi_{\alpha\eta} m_{\alpha\eta \mathbf{k}}, \\
V^{b,f(c)}_\mu(q,k) &= (1 - \delta_{\mu 0}) \cdot \frac{il}{2} N \sum_\alpha j_\alpha M_{\alpha\eta} W^{b,f}_{\alpha\eta}(q,k), \\
V^{b,f(d)}_\mu(q,k) &= N \sum_{\tilde{\alpha}_b} \sum_{k'} \mathrm{Tr} \left[ V^{b(b)}_\mu(q,k') G_{k'+\frac{q}{2}} e^{-i\frac{q\eta}{2}} m_{\alpha\eta \mathbf{k}'} G_{k'-\frac{q}{2}} \right] W^{b,f}_{\alpha\eta}(q,k) \\
&\quad - N \sum_{\tilde{\alpha}_f} \sum_{k'} V^{f(b)}_\mu(q,k') g_{k'+\frac{q}{2}} e^{-i\frac{q\eta}{2}} m_{\alpha\eta \mathbf{k}'} g_{k'-\frac{q}{2}} W^{b,f}_{\alpha\eta}(q,k).
\end{aligned} \tag{5.21}$$

Here, like in Eq. (2.58), in the top (bottom) line of the last formula $m_{7\mathbf{k}}$ is defined as $m^b_{7\mathbf{k}}$ ($m^f_{7\mathbf{k}}$), and

$$W^{b,f}_{\alpha\eta}(q,k) \equiv \sum_{\tilde{\beta}_{b,f}} D_{\eta\alpha\eta'\beta}(q) U_{\beta\eta'}(k,q) \tag{5.22}$$

corresponds to the propagation of AF and its interaction with the $b$-$f$ line on the diagrams Fig. 4(c,d).

Consider finally the interaction with EMF, produced by the current $j_\eta^s(x)$, Eq. (5.16). In $j_\eta^s$, "s" stands for "symmetric". Unlike other terms in (5.13), this current exists in both symmetric and asymmetric phases, so that in the symmetric phase with $M_{\alpha\eta} = 0$ only this term gives a nonvanishing contribution to $j_\eta(x)$. The current $j_\eta^s(x)$ creates the vertex with a vector potential, inserted at the $\phi$-$bf$ interaction vertex. The corresponding segment is presented on Fig. 7. It corresponds to the analytic expression

$$-e A_\eta(-q) F^{b,f}_{k-\frac{q+p}{2}} \sum_{\alpha_{b,f}} V_{\alpha\eta}(k,p,q) D_{\eta\alpha\eta'\alpha'}(-p) F^{b,f}_{k+\frac{q+p}{2}}, \tag{5.23}$$

where the functions $F^{b,f}$ have the same meaning as in Eq. (5.20), and

$$V_{\alpha\eta}(k,p,q) = \frac{il}{2} j_\alpha m_{\alpha\eta} \mathbf{k} e^{i\frac{q_\eta + p_\eta}{2}}. \tag{5.24}$$

We note that just like with the diagrams of Fig. 5(a,b), the diagrams of Fig. 5(c,d) are legitimate, although the diagram of Fig. 1(b) is forbidden.

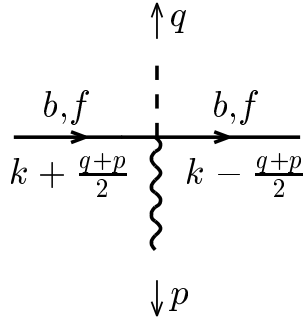

FIG. 7. The vertex of interaction with the EMF, corresponding to the current $j_\eta^s(x)$, Eq. (5.16).

## VI. Ward identities

In this section we will show that the amplitudes of interaction with EMF in our theory satisfy the Ward identities

$$i\omega \Gamma_0(q,k) + d_{-q\eta} \Gamma_\eta(q,k) = \tau_3 G^{(e)}_{k+\frac{q}{2}} - G^{(e)}_{k-\frac{q}{2}} \tau_3. \tag{6.1}$$

Here $\omega$ is the zero component of $q = (\omega, \mathbf{q})$, and we have introduced the amplitudes corresponding to the electron's interaction with EMF

$$\Gamma_0(q,k) \equiv \frac{1}{\beta V} \langle n_q \Psi^{\xi\zeta}_{k-\frac{q}{2}} \Psi^{\xi\zeta\dagger}_{k+\frac{q}{2}} \rangle, \tag{6.2}$$

$$\Gamma_\eta(q,k) \equiv \frac{1}{\beta V} \langle j_\eta(q) \Psi^{\xi\zeta}_{k-\frac{q}{2}} \Psi^{\xi\zeta\dagger}_{k+\frac{q}{2}} \rangle, \tag{6.3}$$

where $\xi$ and $\zeta$ are color indices, $\Psi^{\xi\zeta}_k$ is a Fourier component of the electron field

$$\Psi^{\xi\zeta}_x \equiv \begin{pmatrix} b^\xi_{\uparrow x} f^\zeta_x \\ b^{\xi*}_{\downarrow x} f^{\zeta*}_x \end{pmatrix}, \tag{6.4}$$

and $G^{(e)}_k$ is the electron Green's function:

$$G^{(e)}_k \equiv \frac{1}{\beta V} \langle \Psi^{\xi\zeta}_k \Psi^{\xi\zeta\dagger}_k \rangle. \tag{6.5}$$

The summation over repeating color indices is assumed.

In the rest of this section, we present the path-integral and the diagrammatic derivations of the Ward identities (6.1). It will be clear from these derivations that the Ward identities are satisfied in each order of the $1/N$ expansion separately. Only the Ward identities themselves, and not these derivations, will be used in the forthcoming sections.

## VI.1 Path-integral derivation

We start by introducing the source terms and considering the theory determined by the Lagrangian

$$\mathcal{L}_K(x) = \mathcal{L}(x) - K_x^{\xi\zeta\,\dagger}\,\Psi_x^{\xi\zeta} - \Psi_x^{\xi\zeta\,\dagger}\,K_x^{\xi\zeta}\,, \tag{6.6}$$

where $\mathcal{L}(x)$ is the Lagrangian in (2.11), $\Psi_x^{\xi\zeta}$ is an electron field (6.4), and $K_x^{\xi\zeta}$ is its source:

$$K_x^{\xi\zeta} = \begin{pmatrix} K_{\uparrow x}^{\xi\zeta} \\ K_{\downarrow x}^{\xi\zeta\,*} \end{pmatrix}. \tag{6.7}$$

The summation over repeating color indices is assumed; in the following, we will make these indices implicit. Let $Z_K$ be the partition function, corresponding to the Lagrangian (6.6):

$$Z_K = \int \mathcal{D}(c) e^{-\mathcal{S}_K}\,, \tag{6.8}$$

where $\mathcal{S}_K \equiv \int \mathcal{L}_K(x)\,dx$ and $c_x = b_x, b_x^*, f_x, f_x^*$ is the general notation for the fields, participating in (6.8). We now make in the path integral (6.8) the change of integration variables $b_x \to e^{i\alpha_x \tau_3} b_x$, $b_x^* \to b_x^* e^{-i\alpha_x \tau_3}$, $f_x \to e^{i\alpha_x} f_x$, $f_x^* \to e^{-i\alpha_x} f_x^*$. We have for small $\alpha_x$ the variations $\delta b_x = i\alpha_x \tau_3 b_x$, $\delta \dot{b}_x = i\alpha_x \tau_3 \dot{b}_x + i\dot{\alpha}_x \tau_3 b_x$, etc. The condition that $Z_K$ does not change with the change of integration variables yields

$$\int \mathcal{D}(c)\,\frac{\delta \mathcal{S}_K}{\delta(i\alpha_x)}\,e^{-\mathcal{S}_K} = 0\,. \tag{6.9}$$

We now take into account that the Lagrangian $\mathcal{L}(x)$ depends on the products of fields at the neighbouring points, so it has the structure

$$\mathcal{L}(x) = \mathcal{L}_x(\dot{c}_x, c_x, \{c_{x+\hat{\eta}}, \eta = 1,\ldots,d\})\,. \tag{6.10}$$

We have then:

$$\begin{aligned}
\delta \mathcal{S}_K &= \int dx \Bigg\{ \sum_c \Bigg[ i\alpha_x \left( \left( \frac{\partial \mathcal{L}_x}{\partial c_x} + \sum_\eta \frac{\partial \mathcal{L}_{x-\hat{\eta}}}{\partial c_x} \right) \frac{\delta c_x}{\delta(i\alpha_x)} + \frac{\partial \mathcal{L}_x}{\partial \dot{c}_x} \frac{\delta \dot{c}_x}{\delta(i\alpha_x)} \right) \\
&\qquad + i\dot{\alpha}_x \frac{\partial \mathcal{L}_x}{\partial \dot{c}_x} \frac{\delta \dot{c}_x}{\delta(i\alpha_x)} \Bigg] - 2i\alpha_x \left( K_x^\dagger \tau_3 \Psi_x - \Psi_x^\dagger \tau_3 K_x \right) \Bigg\} \\
&= \int dx\, i\alpha_x \Bigg\{ \sum_c \Bigg[ \frac{\partial \mathcal{L}_x}{\partial c_x} \frac{\delta c_x}{\delta(i\alpha_x)} + \frac{\partial \mathcal{L}_x}{\partial c_{x+\hat{\eta}}} \frac{\delta c_{x+\hat{\eta}}}{\delta(i\alpha_{x+\hat{\eta}})} + \frac{\partial \mathcal{L}_x}{\partial \dot{c}_x} \frac{\delta \dot{c}_x}{\delta(i\alpha_x)} \\
&\qquad + \left( \sum_\eta \frac{\partial \mathcal{L}_{x-\hat{\eta}}}{\partial c_x} \frac{\delta c_x}{\delta(i\alpha_x)} - \frac{\partial \mathcal{L}_x}{\partial c_{x+\hat{\eta}}} \frac{\delta c_{x+\hat{\eta}}}{\delta(i\alpha_{x+\hat{\eta}})} \right) - \frac{\partial}{\partial \tau} \left( \frac{\partial \mathcal{L}_x}{\partial \dot{c}_x} \frac{\delta \dot{c}_x}{\delta(i\alpha_x)} \right) \Bigg] \\
&\qquad - 2\left( K_x^\dagger \tau_3 \Psi_x - \Psi_x^\dagger \tau_3 K_x \right) \Bigg\}.
\end{aligned} \tag{6.11}$$

Now (6.9) gives

$$\langle Y_x + i\partial_\eta^L\,j_\eta^W(x) - \frac{\partial n_x^W}{\partial \tau} + \Psi_x^\dagger \tau_3 K_x - K_x^\dagger \tau_3 \Psi_x \rangle_K = 0\,. \tag{6.12}$$

Here $\langle\cdot\rangle_K$ means the average in the theory with the Lagrangian $\mathcal{L}_K(x)$, $Y_x = \frac{1}{2}\frac{\delta\mathcal{L}_x}{\delta(i\alpha_{\text{const}})}$, where

$$\frac{\delta\mathcal{L}_x}{\delta(i\alpha_{\text{const}})} \equiv \sum_c \left[\frac{\partial\mathcal{L}_x}{\partial c_x}\frac{\delta c_x}{\delta(i\alpha_x)} + \frac{\partial\mathcal{L}_x}{\partial c_{x+\hat{\eta}}}\frac{\delta c_{x+\hat{\eta}}}{\delta(i\alpha_{x+\hat{\eta}})} + \frac{\partial\mathcal{L}_x}{\partial \dot{c}_x}\frac{\delta \dot{c}_x}{\delta(i\alpha_x)}\right] \tag{6.13}$$

is the derivative of the Lagrangian with respect to the constant, $x$-independent phase, and the current $j_\eta^W(x)$ and density $n_x^W$ are given by

$$j_\eta^W(x) \equiv \frac{il}{2}\sum_c \frac{\partial\mathcal{L}_x}{\partial c_{x+\hat{\eta}}}\frac{\delta c_{x+\hat{\eta}}}{\delta(i\alpha_{x+\hat{\eta}})} \tag{6.14}$$

$$n_x^W \equiv \frac{1}{2}\sum_c \frac{\partial\mathcal{L}_x}{\partial \dot{c}_x}\frac{\delta \dot{c}_x}{\delta(i\alpha_x)}. \tag{6.15}$$

It is straightforward to check that for the Lagrangian $\mathcal{L}$ from Eq. (2.11), $j_\eta^W(x)$ and $n_x^W$ defined by Eqs. (6.14) – (6.15) coincide correspondingly with $j_\eta(x)$ and $n_x$.

Since $\mathcal{L}$ is invariant with respect to the phase transformation with a constant phase, the first term in Eq. (6.12) vanishes, and Fourier transformation of this equation yields

$$\langle i\omega n_q + d_{-q\eta}\, j_\eta(q) + \sum_k \left(\Psi^\dagger_{k-\frac{q}{2}}\tau_3 K_{k+\frac{q}{2}} - K^\dagger_{k-\frac{q}{2}}\tau_3 \Psi_{k+\frac{q}{2}}\right)\rangle_K = 0. \tag{6.16}$$

Now differentiating this equation with respect to $K^{\xi\zeta\,\dagger}_{k-\frac{q}{2}}$ and $K^{\xi\zeta}_{k+\frac{q}{2}}$ and setting $K^{\xi\zeta\,\dagger}_x = K^{\xi\zeta}_x = 0$ we obtain the Ward identities (6.1).

## VI.2 Diagrammatic derivation

Here we want to show how our diagram technique satisfies the Ward identities. We first note that the form (6.1) of the Ward identities is not in a straightforward correspondence with our diagram technique. Indeed, as we have just seen, this form corresponds to the Lagrangian (2.11), rather then to the Lagrangian (2.22), which generates our diagram technique. With the Lagrangian (2.22) we obtain a different expression for the current (6.14) and for the $Y$-term (6.13). The current (6.14) with the Lagrangian (2.22) is

$$\begin{aligned}j'_\eta(x) &= j_\eta(x) - \tilde{j}_\eta(x),\\ \tilde{j}_\eta(x) &= \frac{il}{2}\sum_\alpha \tilde{g}_\alpha\left[\phi_{\alpha\eta}M'_{\alpha\eta x} + \phi_{\alpha\eta x}\left(NM_{\alpha\eta} + M'_{\alpha\eta x}\right)\right].\end{aligned} \tag{6.17}$$

The new contribution to the current, $-\tilde{j}_\eta(x)$, may be also obtained via differentiation of the Lagrangian with respect to $-eA_\eta(x)$, if along with the substitution (5.2) we also make in (2.22) the change of integration variables $\phi_{\alpha\eta x} \to \exp[-iel\tilde{g}_\alpha A_\eta(x)/2]\phi_{\alpha\eta x}$. By the equations of motion (4.12) this new current vanishes; however, in our diagrammatic proof of the Ward identities it will be easier for us to keep it in the formulas. The action of $\tilde{j}_\eta(x)$ is described by the same diagrams Fig. 4 – 7 as the action of $j_\eta(x)$. The corresponding analytic expressions are obtained from Eqs. (5.20) – (5.24) (the second equality should be used for $V_\eta^{b,f(b)}(q,k)$) by the simple substitution $j_\alpha \to \tilde{g}_\alpha$; we will mark the corresponding vertex functions with a tilde.

The $Y$-term (6.13) corresponding to the Lagrangian (2.22) is

$$Y(x) = -\frac{2}{il}\sum_\eta \tilde{j}_\eta(x). \tag{6.18}$$

Performing the Fourier transformation of Eq. (6.12) and differentiating it with respect to the sources, we obtain that the amplitudes in our diagram technique must satisfy the equation

$$\tau_3 G^{(e)}_{k+\frac{q}{2}} - G^{(e)}_{k-\frac{q}{2}}\tau_3 = i\omega\Gamma_0(q,k) + d_{-q\eta}\left[\Gamma_\eta(q,k) - \widetilde{\Gamma}_\eta(q,k)\right] - \frac{2}{il}\sum_\eta \widetilde{\Gamma}_\eta(q,k), \tag{6.19}$$

where
$$\widetilde{\Gamma}_\eta(q,k) = \frac{1}{\beta V} \langle \tilde{j}_\eta(q) \Psi^{\xi\zeta}_{k-\frac{q}{2}} \Psi^{\xi\zeta\,\dagger}_{k+\frac{q}{2}} \rangle. \tag{6.20}$$

The Ward identities (6.1) follow from (6.19) immediately, since $\widetilde{\Gamma}_\eta(q,k)$ vanishes by the equations of motion. Using the equality $d_{-q\eta} + 2(il)^{-1} = -\tilde{d}_{-q\eta}$, we now present (6.19) in a form

$$i\omega \Gamma_0(q,k) + d_{-q\eta} \Gamma_\eta(q,k) + \tilde{d}_{-q\eta} \widetilde{\Gamma}_\eta(q,k) = \tau_3 G^{(e)}_{k+\frac{q}{2}} - G^{(e)}_{k-\frac{q}{2}} \tau_3, \tag{6.21}$$

which is convenient for the diagrammatic proof.

In this proof we will use the following important relation, derived in Appendix:

$$i\omega\, V_0^c(q,k) + d_{-q\eta} V_\eta^c(q,k) + \tilde{d}_{-q\eta} \widetilde{V}_\eta^c(q,k) = \begin{cases} \frac{1}{2}\left( G^{-1}_{k-\frac{q}{2}} \tau_3 - \tau_3\, G^{-1}_{k+\frac{q}{2}} \right), & c = b \\ \frac{1}{2}\left( g^{-1}_{k-\frac{q}{2}} - g^{-1}_{k+\frac{q}{2}} \right), & c = f \end{cases} \tag{6.22}$$

It is evident from the derivation of (6.22) in Appendix that the vertex $V_\mu^{(n)}(q)$, Fig. 6, satisfies the equation

$$i\omega\, V_0^{(n)}(q) + d_{-q\eta} V_\eta^{(n)}(q) + \tilde{d}_{-q\eta} \widetilde{V}_\eta^{(n)}(q) = 0. \tag{6.23}$$

Another important relation, which connects the $\phi$-$bf$ interaction vertex (2.48) with the vertex (5.24), is

$$d_{-q\eta} V_{\alpha\eta}(k,p,q) + \tilde{d}_{-q\eta} \widetilde{V}_{\alpha\eta}(k,p,q)$$
$$= \frac{1}{2}\left[ \tilde{\tau}_\alpha U_{\alpha\eta}\left(k+\frac{q}{2},p\right) - U_{\alpha\eta}\left(k-\frac{q}{2},p\right) \tilde{\tau}_\alpha \right], \tag{6.24}$$

where $\tilde{\tau}_\alpha$ are defined after Eq. (A1). To prove this relation, we substitute in the right hand side of (6.24) the definition (2.48) of $U_{\alpha\eta}\left(k \pm \frac{q}{2},p\right)$ and use (A1) to obtain the left hand side.

The factors $1/2$ on the right hand sides of Eqs. (6.22) and (6.24) correspond to the symmetric choice $\kappa_b = \kappa_f = 1/2$ in (5.7); we recall, however, that as we proved in a general form, the different choices of $\kappa_{b,f}$ will give the same results.

We now go on directly to the derivation of the Ward identities (6.21). We will derive them in the usual way [19], using the fact that (with one exeption, which we will discuss later) each diagram of the electron's interaction with EMF may be obtained by the insertion of the EMF line into the diagram of the electron's Green's function. According to the previous section, the line of the scalar potential may be inserted only at the propagators of AF, spinons and monons, and at the vertex of self-interaction of AF, while the line of vector potential may be also inserted at the vertex of interaction of AF with spinons and monons. By the interaction vertex of EMF with spinons and monons we will understand the renormalized vertex of Fig. 4(a), described by $V_\mu^{b,f}(q,k)$. Then the disconnection of EMF line from any diagram for the amplitude of the electron's interaction with EMF results in the possible diagram for the electron's Green's function. On the other hand, if we understood by the interaction vertex of EMF with spinons and monons only the sum of the vertices Fig. 4(b,c), then the disconnection of EMF line from this vertex could result in the forbidden diagram for the Green's function, as is clear from the segment of Fig. 4(d).

Let us now consider a diagram for the electron Green's function (Fig. 8). In order to get the corresponding diagrams for the electron's interaction with EMF, the EMF line may be inserted at each propagator or vertex of this diagram. Let us consider first insertions at the $b$-line on the top of the diagram. For the evaluation of the contribution of each insertion to the left hand side of Eq. (6.21) we use, depending on the position of insertion, either equation (6.24) or the first equation (6.22). Then upon summation over all possible positions where the EMF line can be inserted, we see that the terms in the sum cancel pairwise; only two terms coming from the insertions of EMF line to the extreme left and the extreme right positions on the $b$-line escape the cancellation and give precisely a half of the right hand side of Eq. (6.21). Another half comes in the same way from the insertions at the $f$-line on the bottom of the diagram. Then, the insertions at the internal $bf$-loops lead simply to the shift of the integration variable in the corresponding momentum integrals and give, therefore, zero contribution. By (6.23), the insertions at the AF propagators and at the vertices of self-interaction of AF also give zero contribution. Finally we note that there exists an exception from the rule of the correspondence between the diagrams for the electron Green's function and the electron's interaction with EMF. Namely, the latter diagrams may include segments of Fig. 5, and the disconnection of EMF line from these segments leads to the forbidden segment of Fig. 1(b). However, this exception does not spoil our proof, because the diagrams of Fig. 5, being the insertions at the internal loop, give zero contribution to the left hand side of Eq. (6.21), as was just explained.

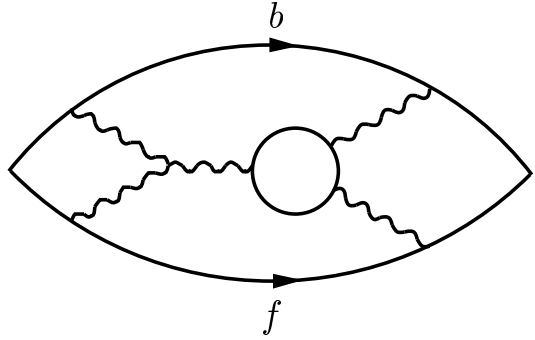

FIG. 8. A typical diagram for the electron Green's function. Insertion of EMF line at each propagator or vertex of this diagram produces the possible diagram for the electron's interaction with EMF.

## VII. Currents correlation singularities and superconducting state

The reason for deriving the Ward identities is that they provide us with a simple way to demonstrate gauge invariance of the electric current and to detect the superconducting state. In the first order in EMF, the Fourier component of the electric current is

$$\langle j_\eta^e(q)\rangle = \left[K_{\eta\eta'}^{(e)}(q) - \widetilde{\rho}_\eta\,\delta_{\eta\eta'}\right] A_{\eta'}(q) - K_{\eta 0}^{(e)}(q)\,\varphi(q), \tag{7.1}$$

where the electromagnetic kernels $K_{\eta\mu}^{(e)}(q)$ are

$$\begin{aligned}
K_{\eta\eta'}^{(e)}(q) &\equiv \frac{e^2}{\beta V}\,\langle j_\eta(q)\,j_{\eta'}(-q)\rangle, \\
K_{\eta 0}^{(e)}(q) &\equiv \frac{e^2}{\beta V}\,\langle j_\eta(q)\,n_{-q}\rangle,
\end{aligned} \tag{7.2}$$

and

$$\widetilde{\rho}_\eta \equiv \frac{e^2 l^2 t_\eta}{N}\left\langle b^\dagger_{\sigma\mathbf{r}+\hat{\eta}}b_{\sigma\mathbf{r}}f^\dagger_{\mathbf{r}+\hat{\eta}}f_{\mathbf{r}} + b^\dagger_{\sigma\mathbf{r}}b_{\sigma\mathbf{r}+\hat{\eta}}f^\dagger_{\mathbf{r}}f_{\mathbf{r}+\hat{\eta}}\right\rangle. \tag{7.3}$$

In terms of electron fields $\Psi_k^{\xi\zeta}$, the Fourier component of electron current (5.3) is

$$j_\eta(q) = \frac{2lt_\eta}{N}\,e^{i\frac{q_\eta}{2}}\sum_k \Psi^{\xi\zeta\,\dagger}_{k-\frac{q}{2}}\Psi^{\xi\zeta}_{k+\frac{q}{2}}\,\sin k_\eta. \tag{7.4}$$

Therefore, the kernels may be expressed through amplitudes (6.2) and (6.3) as

$$\begin{aligned}
K_{\eta\eta'}^{(e)}(q) &= -\frac{2e^2 l\,t_\eta}{N}\,e^{i\frac{q_\eta}{2}}\sum_k \operatorname{Tr}\Gamma_{\eta'}(-q,k)\cdot\sin k_\eta, \\
K_{\eta 0}^{(e)}(q) &= -\frac{2e^2 l\,t_\eta}{N}\,e^{i\frac{q_\eta}{2}}\sum_k \operatorname{Tr}\Gamma_0(-q,k)\cdot\sin k_\eta.
\end{aligned} \tag{7.5}$$

We have, also,

$$\widetilde{\rho}_\eta = -\frac{2e^2 l^2 t_\eta}{N} \sum_k \text{Tr}\left[G_k^{(e)} \tau_3\right] \cos k_\eta . \qquad (7.6)$$

From the Ward identities (6.1) we have then

$$\begin{aligned} i\omega K_{\eta 0}^{(e)}(q) - d_{q\eta'} K_{\eta\eta'}^{(e)}(q) &= \frac{2e^2 l t_\eta}{N} e^{i\frac{q_\eta}{2}} \sum_k \text{Tr}\left[\tau_3 G_{k-\frac{q}{2}}^{(e)} - G_{k+\frac{q}{2}}^{(e)} \tau_3\right] \sin k_\eta \\ &= -\widetilde{\rho}_\eta d_{q\eta} , \end{aligned} \qquad (7.7)$$

where we have made a shift in the summation variable and have used (7.6). Now from (7.1) we see that our current is gauge-invariant: the pure gauge external field $(\varphi(q), A_\eta(q)) \sim (i\omega, d_{q\eta})$ produces no current.

In an external magnetic field, described in a gauge $\varphi = 0$ by a vector potential $A_\eta(\mathbf{r}) = A_\eta(\mathbf{q}) e^{i\mathbf{q}\mathbf{r}}$, the Fourier component of a current is given by the first two terms in (7.1), where $q = (0, \mathbf{q})$. By (7.7), the kernel there satisfies the equation

$$d_{q\eta'} K_{\eta\eta'}^{(e)}(\mathbf{q}) = \widetilde{\rho}_\eta d_{q\eta} . \qquad (7.8)$$

Now differentiating this equation over $q_{\eta'}$ yields

$$\lim_{\mathbf{q}\to 0}\left[K_{\eta\eta'}^{(e)}(\mathbf{q}) + q_{\eta''} \frac{\partial K_{\eta\eta''}^{(e)}(\mathbf{q})}{\partial q_{\eta'}}\right] = \widetilde{\rho}_\eta \delta_{\eta\eta'} . \qquad (7.9)$$

Therefore, if $K_{\eta\eta'}^{(e)}(\mathbf{q})$ is a regular function of $\mathbf{q}$ at $\mathbf{q} = 0$, then (7.9) gives $K_{\eta\eta'}^{(e)}(0) = \widetilde{\rho}_\eta \delta_{\eta\eta'}$, and the diamagnetic and paramagnetic contributions to (7.1) cancel each other near $\mathbf{q} = 0$; the system is in the normal state. If, on the other hand, $K_{\eta\eta'}^{(e)}(\mathbf{q})$ is singular, so that the second term in (7.9) has a nonzero limit at $\mathbf{q} = 0$, then this cancellation does not happen, and in the external magnetic field the current with momentum $\mathbf{q} \approx 0$ does not vanish; the system is in the superconducting state.

The required singularity in the current-current correlation function (in the BCS theory this singularity has a simple form, $q_\eta q_{\eta'}/q^2$) is known to exist in the superfluid phase of the system of interacting bosons. Consequently, if these bosons are charged, then this system exhibits superconductivity [20]. For the case of the usual 3D superfluidity with nonzero expectation value of the boson field, this singularity was considered in [21]; the singularity in the 2D superfluid state of the Berezinskii–Kosterlitz–Thouless (BKT) type was discussed in [22].

The spinon subsystem in the $t$-$J$ model may in principle exhibit both types of superfluidity. The study of the BKT superfluidity requires application of methods essentially different from those used in this paper [23], and so here we only consider in detail the case of 3D superfluidity which corresponds to $v_{\mathbf{r}} \neq 0$. Let also $B_\eta$ and $F_\eta$ have nonzero values, so that the vertices $V_\eta^{b,f(b)}(q,k)$, Eq. (5.21), do not vanish. Using Eq. (5.19) we obtain that on the MF level the current-current correlation function $K_{\eta\eta'}^{(e)}(q)$ is given by the two diagrams, presented in Fig. 9, both of order $N$ (or of order 1 per color). We need to evaluate these diagrams for $q = (0, \mathbf{q})$. The polarization matrix $\Pi'(\mathbf{q})$ (which is present in these diagrams explicitly, as well as implicitly, through the propagator of AF) includes the condensate contribution, described by the diagrams of Fig. 10. We will denote this contribution $\Pi'^v(\mathbf{q})$. The relevant quantity is the current-current matrix element of $\Pi'^v(\mathbf{q})$:

$$\Pi'^v_{j^b_\eta j^b_{\eta'}}(\mathbf{q}) = -l^2 t_\eta t_{\eta'} M_{f\eta} M_{f\eta'} \sum_{\alpha_b \alpha'_b} j_\alpha j_{\alpha'} \Pi'^v_{\eta\alpha\eta'\alpha'}(\mathbf{q})$$

$$= 4Nl^2 t_\eta t_{\eta'} M_{f\eta} M_{f\eta'} e^{\frac{i}{2}(q_\eta - q_{\eta'})} v^\dagger_{\mathbf{k}_c} \tau_3 \bigg[ G_{\mathbf{k}_c - \mathbf{q}} \sin\left(k_{c\eta} - \frac{q_\eta}{2}\right) \sin\left(k_{c\eta'} - \frac{q_{\eta'}}{2}\right)$$

$$+ G_{\mathbf{k}_c + \mathbf{q}} \sin\left(k_{c\eta} + \frac{q_\eta}{2}\right) \sin\left(k_{c\eta'} + \frac{q_{\eta'}}{2}\right) \bigg] \tau_3 v_{\mathbf{k}_c} + (\mathbf{k}_c \longrightarrow -\mathbf{k}_c), \quad (7.10)$$

where we have used Eq. (5.14). The Green's functions $G_{\mathbf{k}_c \pm \mathbf{q}}$ in Eq. (7.10) are given by Eq. (2.41), where we have to set $k = (0, \mathbf{k}_c \pm \mathbf{q})$. The denominator of Eq. (2.41) is then equal to $\omega^2_{\mathbf{k}_c \pm \mathbf{q}}$, which is proportional to $q^2$ for small $q$. Now the straightforward evaluation of Eq. (7.10) shows that for small $q$ the current-current matrix element of $\Pi'^v(\mathbf{q})$ is proportional to $q_\eta q_{\eta'}/q^2$. Therefore, we can expect that the diagrams with the AF propagator in the intermediate state give the required singular contribution to the current-current correlation function.

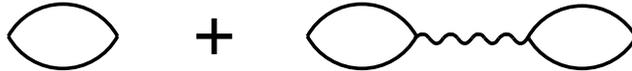

FIG. 9. MF diagrams for the electromagnetic kernel $K_{\eta\eta'}^{(e)}(q)$.

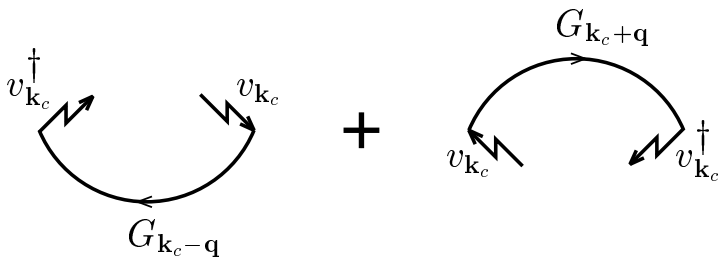

FIG. 10. Diagrammatic representation for the condensate contribution to the polarization matrix, $\Pi'^v(\mathbf{q})$.

However, this does not happen for the diagrams of Fig. 9. Indeed, on the MF level our problem is essentially similar to the one, considered by Ioffe and Larkin [24] and by Lee and Nagaosa [25]. As was first shown by the functional methods in Ref. [24], when the fermion subsystem is in the normal state, the singularities in the diagrams of Fig. 9 must cancel out, so that the whole electron system is also in the normal state. It is instructive to obtain this cancellation in the framework of our approach.

We first introduce the susceptibility matrix $K_{\eta\alpha\eta'\alpha'}(q)$, corresponding to the diagrams of Fig. 9, by

$$K(q) \equiv \Pi'(q) + \Pi'(q) D(q) \Pi'(q). \quad (7.11)$$

This matrix describes all kinetic properties of the system on the MF level. For example, the MF electromagnetic kernel $K^{(e)\mathrm{MF}}_{\eta\eta'}(q)$ is a current-current matrix element of $K(q)$:

$$K^{(e)\mathrm{MF}}_{\eta\eta'}(q) = -e^2 l^2 \sum_{\alpha\alpha'} j_\alpha \phi_{\alpha\eta} K_{\eta\alpha\eta'\alpha'}(q) j_{\alpha'} \phi_{\alpha'\eta'}, \quad (7.12)$$

where we have used Eqs. (5.14) and (5.19). Using (2.57), (2.59), and the Fourier representation $\Pi^{(b)}_{\eta\alpha\eta'\alpha'}(q) = \Pi^{(0)}_{\eta\alpha\alpha'}\delta_{\eta\eta'} + \Pi^{g(0)}_{\eta\alpha\eta'\alpha'}(q)$ of the bare polarization matrix $\Pi^{(b)}_{\eta\alpha\eta'\alpha'}(x - x')$, Eq. (4.4), we can present (7.11) as a sum over all possible chains, built up from the RPA polarization bubbles:

$$K(q) = \Pi'(q) \sum_{n=0}^{\infty} \left[ \Pi^{(b)-1}(q) \Pi'(q) \right]^n. \quad (7.13)$$

Then we separate the boson and fermion contributions to the polarization matrix $\Pi'(q)$:

$$\Pi'(q) = \Pi'^b(q) + \Pi'^f(q), \quad (7.14)$$

where $\Pi'^b$ ($\Pi'^f$) is described by the boson (fermion) RPA polarization bubble. The nonzero matrix elements of $\Pi'^{b,f}(q)$ are given, respectively, by the first and second lines of Eq. (2.58). Now for $q = (0, \mathbf{q})$ with small $\mathbf{q}$ we can present (7.13) in a usefull form:

$$K(\mathbf{q}) = \Pi'^f(\mathbf{q}) + \left[1 + \Pi'^f(\mathbf{q})\frac{h}{N}\right]\widetilde{\Pi}'^b(\mathbf{q})$$
$$\times \sum_{n=0}^{\infty}\left[\Pi^{(b)^{-1}}(\mathbf{q})\,\Pi'^f(\mathbf{q})\,\Pi^{(b)^{-1}}(\mathbf{q})\,\widetilde{\Pi}'^b(\mathbf{q})\right]^n \left[1 + \frac{h}{N}\Pi'^f(\mathbf{q})\right], \quad (7.15)$$

where

$$\widetilde{\Pi}'^b(\mathbf{q}) = \Pi'^b(\mathbf{q})\sum_{n=0}^{\infty}\left[\Pi^{(b)^{-1}}(\mathbf{q})\,\Pi'^b(\mathbf{q})\right]^n. \quad (7.16)$$

In this representation we have taken into account that in each chain (7.13) a fermion bubble $\Pi'^f$ may be coupled only to a boson bubble $\Pi'^b$, while boson bubbles may be coupled to each other due to the presence of the superexchange term in the Hamiltonian. We have also taken into account that the terms in $\Pi^{(g)0}(\mathbf{q})$, which could give nonzero contribution to the matrix product of $\Pi^{(b)^{-1}}(\mathbf{q})$ and $\Pi'^f(\mathbf{q})$, vanish for $\mathbf{q} = 0$. This allowed us to replace $\Pi^{(b)^{-1}}(\mathbf{q})$ by $\Pi^{(0)^{-1}} = h/N$ in the leftmost and the rightmost factors of the second term in Eq. (7.15).

It is convenient to use in (7.15) the basis set $j_\eta^{b,f}(x)$, $\widetilde{n}_\eta^{b,f}(x)$ instead of the set $M_{3\eta x}, \ldots, M_{6\eta x}$, Eq. (2.7). Here

$$\begin{aligned}\widetilde{n}_\eta^b(x) &\equiv l^2 \sum_{\alpha=3,4} \phi_{\alpha\eta} M_{\alpha\eta x}, \\ \widetilde{n}_\eta^f(x) &\equiv l^2 \sum_{\alpha_f} \phi_{\alpha\eta} M_{\alpha\eta x},\end{aligned} \quad (7.17)$$

while $j_\eta^{b,f}(x)$ are given by Eq. (5.14) which we, using Eq. (2.40), rewrite as

$$j_\eta^{b,f}(x) = il \sum_{\alpha_{b,f}} j_\alpha \phi_{\alpha\eta} M_{\alpha\eta x}. \quad (7.18)$$

With the definitions (7.17) and (7.18) our conclusions below will remain correct even when the parameters $M_{\alpha\eta}$, $\alpha = 3, \ldots, 6$ have nonzero imaginary parts.

Using Eqs. (7.17), (7.18), and (2.40), the hopping term $H_t$ in Eq. (2.5) may be written as

$$H_t = \frac{1}{N\widetilde{n}_\eta} \sum_{\mathbf{r}} \left[j_\eta^b(\mathbf{r})j_\eta^f(\mathbf{r}) - \frac{1}{l^2}\widetilde{n}_\eta^b(\mathbf{r})\widetilde{n}_\eta^f(\mathbf{r})\right], \quad (7.19)$$

where we made the $1 \to N$ generalization and denoted as $\widetilde{n}_\eta = 2l^2 t_\eta M_{4\eta} M_{6\eta}$ the common MF expectation value of the operators $\widetilde{n}_\eta^{b,f}(x)$ per color. Therefore, the matrix $h_\eta$ does not couple $j_\eta^{b,f}(x)$ with $\widetilde{n}_\eta^{b,f}(x)$, while the current-current matrix elements of $h_\eta$ are $h_{\eta j^b j^f} = h_{\eta j^f j^b} = -1/\widetilde{n}_\eta$. On the other hand, following the same route as in the derivation of Eq. (7.8), with the help of Eqs. (A1) and (A7) we obtain for the matrix elements of matrix $\Pi'^f(q)$

$$\begin{aligned}\Pi'^f_{j_\eta^f j_{\eta'}^f}(q) &= \frac{1}{\beta V}\langle j_\eta^f(q)\, j_{\eta'}^f(-q)\rangle_{\mathrm{MF}}, \\ \Pi'^f_{j_\eta^f \widetilde{n}_{\eta'}^f}(q) &= \frac{1}{\beta V}\langle j_\eta^f(q)\, \widetilde{n}_{\eta'}^f(-q)\rangle_{\mathrm{MF}}\end{aligned} \quad (7.20)$$

the following relations:

$$\begin{aligned}d_{q\eta'}\Pi'^f_{j_\eta^f j_{\eta'}^f}(\mathbf{q}) &= N\widetilde{n}_\eta d_{q\eta}, \\ d_{q\eta'}\Pi'^f_{j_\eta^f \widetilde{n}_{\eta'}^f}(\mathbf{q}) &= 0.\end{aligned} \quad (7.21)$$

Since $\Pi'^f(\mathbf{q})$ is a regular function of $\mathbf{q}$ at $\mathbf{q} = 0$, Eq. (7.21) means simply that $\Pi'^f_{j^f_\eta j^f_{\eta'}}(0) = N\widetilde{n}_\eta \delta_{\eta\eta'}$ and $\Pi'^f_{j^f_\eta \widetilde{n}^f_{\eta'}}(0) = 0$. It is clear now, that the second term in Eq. (7.15) does not contribute to the MF electromagnetic kernel (7.12) at $\mathbf{q} = 0$, while the contribution of the first term is equal to

$$e^2 \Pi'^f_{j^f_\eta j^f_{\eta'}}(0) = \widetilde{\rho}^{\mathrm{MF}}_\eta \delta_{\eta\eta'}, \tag{7.22}$$

where $\widetilde{\rho}^{\mathrm{MF}}_\eta = e^2 N \widetilde{n}_\eta$ is the MF expectation value of $\widetilde{\rho}_\eta$, Eq. (7.3). Therefore, in agreement with Refs. [24, 25] on the MF level the system is in the normal state [26].

We note however, that this derivation, as well as the derivation of Ref. [25] and the functional derivation of Ref. [24], is valid *only* on the MF level. The cancellation of the singular contributions to the MF electromagnetic kernel occurs between the diagrams, which differ only by one fermion RPA polarization bubble. Consider the higher order diagrams for the electromagnetic kernel, which have the AF propagator in the intermediate state. Among these diagrams one can easily find those, which can not be cancelled in this way, for example the diagrams which use the term $j^s_\eta(x)$, Eq. (5.16), for the current vertices in $K^{(e)}_{\eta\eta'}(q)$. On the other hand, there are no reasons for a cancellation between these diagrams, and so they produce a nonzero singular contribution to the electromagnetic kernel. Therefore, taking into account the corrections of higher orders in $1/N$, we conclude that in the phase with nonzero $v_\mathbf{r}$, $B_\eta$, $F_\eta$ the model exhibits superconductivity even when neither monons nor electrons are paired.

## VIII. Discussion

In this paper we have presented the theory of the $t$-$J$ model in the presence of the external EMF. We have developed the $1/N$ expansion for this model and shown that all the general properties of the theory are satisfied in every order of the $1/N$ expansion separately. We have also shown that superconductivity may exist in this model even in the absence of the electron pairing. Can our approach give a unified picture of the properties of the copper oxides? Let us consider the properties of different phases of the $t$-$J$ model and the possible form of its MF phase diagram. The phases are classified first by the values of the order parameters $\mathcal{A}_\eta$, $B_\eta$ and $F_\eta$, and second by whether the spinon subsystem is in the normal or superfluid state. As was discussed in Sec. II, the parameters $B_\eta$ and $F_\eta$ obtain nonzero values simultaneously. We will label different phases by writing their nonzero order parameters; the symmetric phase will be denoted as $S$.

We must now distinguish between the three- and two-dimensional phases. In the three-dimensional phases all three components of the order parameters $\mathcal{A}_\eta$ and(or) $B_\eta$, $F_\eta$ are different from zero. In this case, if the spinon subsystem is in the superfluid state, it means the real Bose-condensation of spinons with nonzero order parameter $v$. We can have, therefore, the following asymmetric three-dimensional phases: $\mathcal{A}$, $BF$, $\mathcal{A}BF$, $v\mathcal{A}$, $vBF$, and $v\mathcal{A}BF$. One can see from Eq. (2.42) for $\omega_\mathbf{k}$, that in the $v\mathcal{A}$-phase Bose-condensation happens in the state with momentum $k_{c\eta} = \pi/2$; in the $vBF$-phase, with momentum $k_{c\eta} = 0$; and in the $v\mathcal{A}BF$-phase, with momentum $0 < k_{c\eta} < \pi/2$ [11]. As was shown in [12], this Bose-condensation leads to the emergence of the long-range spin order, characterized by a presence of nonzero magnetization with the wave vector $2\mathbf{k}_c$: $\langle \mathbf{S}_{2\mathbf{k}_c} \rangle \neq 0$. We have, therefore, ferromagnetic long-range order (LRO) in the $vBF$-phase, Néel order in the $v\mathcal{A}$-phase, and intermediate ("spiral") LRO in the $v\mathcal{A}BF$-phase [11].

In the two-dimensional phases only two components of $\mathcal{A}_\eta$ and(or) $B_\eta$, $F_\eta$ are different from zero; we will mark these phases with the index "2". According to Hohenberg [27], for nonzero temperature in these phases the spinon field cannot have nonzero expectation value; the superfluid state of the spinon subsystem there is the Berezinskii–Kosterlitz–Thouless state [28]. We will label these phases by the superscript "$SF$" for "superfluid". The two-dimensional asymmetric phases therefore will have such labels as $\mathcal{A}BF_2$, $\mathcal{A}BF_2^{SF}$, etc.

Consider now the properties of the different phases. In the $S$-phase all order parameters are equal to zero. The gauge symmetry is not broken, and the model is in the confining state. Only gauge-invariant objects, like the electron, can propagate, while the propagation of the separated spinons and monons is not possible. On the MF level this is manifested by the momentum-independence of their Green's functions. The currents $j^{b,f,\phi}_\eta(x)$, Eqs. (5.14) – (5.15) vanish, so on the MF level the $S$-phase is insulating. There is also no magnetic order in it.

In the phases $\mathcal{A}_2$ and $\mathcal{A}$ the model exhibits short- and in the phase $v\mathcal{A}$, long-range magnetic order. Like in the $S$-phase, the currents $j^{b,f,\phi}_\eta(x)$ here vanish, so on the MF level these phases are insulating. The physics of these phases is determined by the presence of the gapless modes – $\tilde{a}_{\eta x}$ and, in the $v\mathcal{A}$-phase, spin waves. For the case of half-filling these phases have been treated in [12, 29], and we expect that doping does not change essentially their behavior.

In the phases with nonzero $B$ and $F$ the currents $j^{b,f,\phi}_\eta(x)$ do not vanish. In the $BF$- and $BF_2$-phases the model exhibits the normal conductivity. Here the physics is determined by the gapless gauge field $a_{\eta x}$. The fluctuations of the other parameters of AF $\phi_{\alpha\eta x}$, $\alpha = 3, \ldots, 6$ have gaps. Therefore, their contribution to the AF propagator $D(q)$ in the intermediate state in (7.11) may be neglected compared to the contribution of $a_{\eta x}$, which leads to the Ioffe-Larkin composition rule [18, 24]. We also expect, that other results of [18], where the similar phase was treated in the slave-boson formulation, will be applicable here, which in the first place concerns the linear temperature dependence of the resistivity.

The phases $\mathcal{A}BF$ and $\mathcal{A}BF_2$ are conducting with a short-range magnetic order. As was discussed in Sec. III, the gauge fields $\tilde{a}_{\eta x}$ and $a_{\eta x}$ here acquire gaps due to the Anderson-Higgs mechanism. The spin excitations here also have a gap; these are spin-gap phases. The gap in the excitation spectrum of $a_{\eta x}$ destroys the mechanism which produces the linear temperature dependence of the resistivity [18]. From the experimental point of view, the correlation between the emergence of the spin-gap and the deviation of the temperature dependence of the resistivity from the linear law was recently discussed in [30].

As was shown in the previous section, the phases $vBF$ and $v\mathcal{A}BF$ are superconducting. In the corresponding two-dimensional phases $BF_2^{SF}$ and $\mathcal{A}BF_2^{SF}$ the vertices $V_\eta^{b,f(b)}(q,k)$, Eq. (5.21), do not vanish. Then, although we have not proven it, we also expect (see Ref. [22]), that the current-current correlation function in these phases has a desired zero-momentum singularity. Consequently, these phases are also superconducting. The superconductivity exists in all these two- and three-dimensional phases without monon or electron pairing. We note here the essential difference between two-dimensional superconducting phases $\mathcal{A}BF_2^{SF}$ and $BF_2^{SF}$, and their three-dimensional counterparts $v\mathcal{A}BF$ and $vBF$: in the two-dimensional case all collective modes have gaps, while in the three-dimensional case spin waves are gapless. This difference in the spectra must manifest itself in the neutron scattering.

The possible schematic forms of the MF phase diagrams for the three- and two-dimensional scenarios are presented in Fig. 11(a,b). We note that the very existence of the phases shown there, and their exact bounds may be established, of course, only by the numerical solution of the self-consistency equations (2.37) – (2.39). The general structure of the phase diagram, however, is dictated by a simple condition of continuous dependence of the MF order parameters from the temperature and the hole density. This condition leaves a lot of freedom in placing the phases on the phase diagram. Therefore, a lot of different *types* of phase diagrams are possible. Schematically, the MF phase diagram for the three-dimensional scenario may look like that shown in Fig. 11(a). The three lines on the diagram, which may be called the $v$-line, the $\mathcal{A}$-line, and the $BF$-line, show the boundaries of the phases with nonzero values of the corresponding order parameters. The properties of these phases were considered above. The different phase diagrams may be obtained by moving the lines on Fig. 11(a) with respect to each other. The interesting feature of this phase diagram is the existence of the direct antiferromagnetic-superconductor transition. Experimentally, such a transition was observed in the electron-doped compounds $Nd_{2-x}Ce_xCuO_{4-y}$ and $Pr_{2-x}Ce_xCuO_{4-y}$ [31], so we expect that these compounds are described by the phase diagram of Fig. 11(a). According to the discussion above, the superconducting phases in these materials have magnetic LRO and gapless spin excitations. It would be interesting to check these predictions experimentally.

In this work we did not consider in detail the BKT-transition in the two-dimensional $t$-$J$ model. Nevertheless, we expect that the schematic phase diagram for the two-dimensional scenario might look like that shown in Fig. 11(b). In addition to the lines presented on Fig. 11(a), we have here $\mathcal{A}_2$- and $SF$-lines. Again, by moving the lines on Fig. 11(b), the alternative phase diagrams may be obtained. Thus, the phase diagram of Fig. 11(b) contains a number of possible versions of the phase locations. Now the comparison of this phase diagram with the experimental phase diagrams of the hole-doped high-$T_c$ superconductors $La_{2-x}Sr_xCuO_4$, $YBa_2Cu_3O_{7-\delta}$, etc. suggests that each of these superconductors may be described by one of the possibilities provided by Fig. 11(b); we expect therefore that the hole-doped high-$T_c$ superconductors are two-dimensional in the sense discussed above. This conclusion agrees with the recent experimental results of Shibauchi *et al.* [32], who studied the anisotropic penetration depth in $La_{2-x}Sr_xCuO_4$ crystals. The authors of [32] conclude that the electronic properties of these crystals are essentially two-dimensional. In connection with that, it would be interesting to obtain similar experimental data for $Nd_{2-x}Ce_xCuO_{4-y}$ and $Pr_{2-x}Ce_xCuO_{4-y}$ crystals and to check whether they exhibit three-dimensional behavior as predicted by our approach.

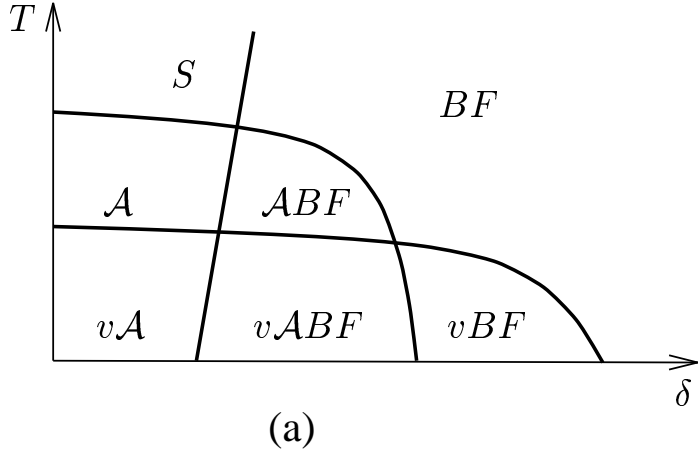

(a)

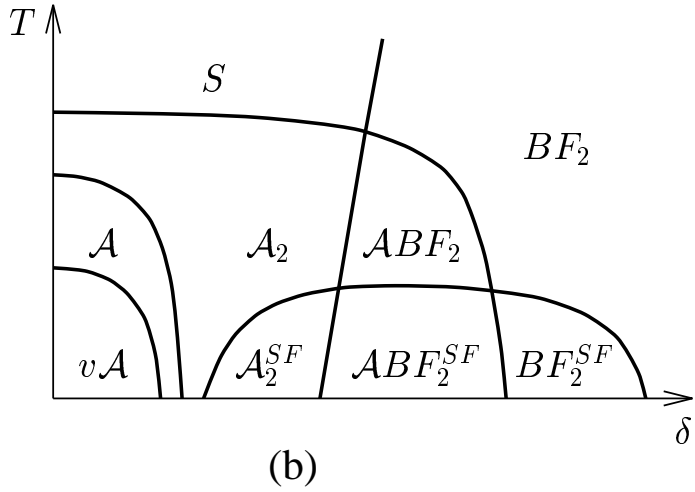

(b)

FIG. 11. Schematic MF phase diagrams. See discussion at the end of Sec. VIII.

Finally, we would like to note that our approach contains in part the idea of charged bosons, which are responsible for the electromagnetic properties of the high-$T_c$ superconductors. The same idea is the basic one for the charged Bose-liquid model of the copper oxides [33]. The authors of [33] state that the unusual properties of the high-temperature superconductors in the superconducting as well as in the normal state are compatible, correspondingly, to those of a degenerate and nondegenerate state of a charged Bose-liquid. They claim that independently of the particular nature of charged bosons, the phenomenological concept of a charged Bose-liquid provides a good description of many experimentally observed properties of the copper oxides. This allows us to hope that our approach will be also successful in the interpretation of experiments.

## Acknowledgments

I am grateful to Boris Altshuler, Assa Auerbach, Michael El-Batanouny, Ganpathy Murthy, Xiao-Gang Wen, and Charles Willis for many useful discussions and valuable comments. I would also like to acknowledge the hospitality of the Institute for Theoretical Physics at Technion, Israel, where a part of this work was done. The work has been supported by the National Science Foundation, Grant No. NSF-DMR-9213884, Department of Energy, Grant No. DE-FG02-91ER45441, and a grant from the US-Israel Binational Science Foundation.

## Appendix: The derivation of (3.12) and (6.22)

We start these derivations with the observation that the coefficients $m_{\alpha\eta\mathbf{k}}$ satisfy the identities

$$\epsilon_\alpha \left( \tilde{\tau}_\alpha m_{\alpha\eta\mathbf{k}-\mathbf{q}/2} - m_{\alpha\eta\mathbf{k}+\mathbf{q}/2} \tilde{\tau}_\alpha \right) = i l e^{-i\frac{q\eta}{2}} \left( \tilde{d}_{q\eta} \tilde{g}_\alpha + d_{q\eta} g_\alpha \right) m_{\alpha\eta\mathbf{k}}, \tag{A1}$$

where for $\alpha \in \widetilde{\mathcal{N}}_b$, $\epsilon_\alpha = 1$ and $\tilde{\tau}_\alpha = \tau_3$, while for $\alpha \in \widetilde{\mathcal{N}}_f$, $\epsilon_\alpha = -1$ and $\tilde{\tau}_\alpha = 1$. We will also use the notations $\tilde{\tau}_b \equiv \tau_3$, $\tilde{\tau}_f \equiv 1$, and $\epsilon_{b,f} \equiv \pm 1$. From (2.35), (2.36), and (A1) we obtain

$$\epsilon_c \left[ \tilde{\tau}_c \left( F^c_{k-\frac{q}{2}} \right)^{-1} - \left( F^c_{k+\frac{q}{2}} \right)^{-1} \tilde{\tau}_c \right] \\ = i\omega \epsilon_c - i l e^{-i\frac{q\eta'}{2}} \sum_{\alpha'_c} \left( \tilde{d}_{q\eta'} \tilde{g}_{\alpha'} + d_{q\eta'} g_{\alpha'} \right) \phi_{\alpha'\eta'} m_{\alpha'\eta'\mathbf{k}}, \tag{A2}$$

where $c$ runs over $b, f$ and the functions $F^c_p$ were defined after Eq. (5.20). In this Appendix we will make the trace over Nambu indices implicit; the sign of Tr will be assumed to be included in the corresponding momenta summation. We will also use here the formulation without the ordering field, so we have $v_{\mathbf{r}} = 0$ in all phases.

Now to prove (3.12) we put in Eq. (A2) $\omega = 0$, multiply it by

$$N e^{i\frac{q\eta''}{2}} F^c_{k-\frac{q}{2}} m_{\alpha''\eta''\mathbf{k}} F^c_{k+\frac{q}{2}} \tag{A3}$$

with $\alpha'' \in \widetilde{\mathcal{N}}_c$, and take a sum over $k$. There is no summation over $\eta''$ in Eq. (A3) and below. Shifting a summation variable, we obtain then for the left hand side of (A2)

$$N e^{i\frac{q\eta''}{2}} \epsilon_c \sum_k \left( \tilde{\tau}_{\alpha''} m_{\alpha''\eta''\mathbf{k}-\mathbf{q}/2} - m_{\alpha''\eta''\mathbf{k}+\mathbf{q}/2} \tilde{\tau}_{\alpha''} \right) F^c_k = i l \epsilon_c N \left( \tilde{d}_{q\eta''} \tilde{g}_{\alpha''} + d_{q\eta''} g_{\alpha''} \right) M_{\alpha''\eta''}, \tag{A4}$$

where we have used (A1) and the self-consistensy equations (2.37) – (2.39); while for the right hand side of (A2) we obtain

$$- i l \epsilon_c \sum_{\alpha'} \Pi'_{\eta''\alpha''\eta'\alpha'}(0, \mathbf{q}) \phi^{(0)}_{\alpha'\eta'\mathbf{q}}. \tag{A5}$$

Rewriting (A4) as

$$i l \epsilon_c \sum_{\alpha'} \left( \tilde{d}_{q\eta''} \tilde{g}_{\alpha''} + d_{q\eta''} g_{\alpha''} \right) \Pi^{(0)}_{\eta''\alpha''\alpha'} \phi_{\alpha'\eta''} = -i l \epsilon_c \sum_{\alpha'} \Pi^{(0)}_{\eta''\alpha''\alpha'} \phi^{(0)}_{\alpha'\eta''\mathbf{q}} \tag{A6}$$

and comparing it with the result for the right hand side, we obtain (3.12).

To prove (6.22), we make in (A2) the change $q \to -q$, multiply it by $\epsilon_c/2$, and use the relations $\epsilon_\alpha \tilde{g}_\alpha = \tilde{g}_\alpha$ and $\epsilon_\alpha g_\alpha = j_\alpha$ to present (A2) in the form

$$i\omega V^{c(b)}_0(q, k) + d_{-q\eta} V^{c(b)}_\eta(q, k) + \tilde{d}_{-q\eta} \widetilde{V}^{c(b)}_\eta(q, k) \\ = \frac{1}{2} \left[ \left( F^c_{k-\frac{q}{2}} \right)^{-1} \tilde{\tau}_c - \tilde{\tau}_c \left( F^c_{k+\frac{q}{2}} \right)^{-1} \right], \tag{A7}$$

which is a contribution of the diagram Fig. 4(b) to Eq. (6.22). Comparing the right hand sides of (A7) and (6.22), we conclude that Eq. (6.22) will be proved if the contributions of the diagrams of Figs. 4(c) and 4(d) to Eq. (6.22) cancel each other.

Let us show that this cancellation does really happen. The contribution of the diagram Fig. 4(d) to Eq. (6.22) is

$$i\omega V_0^{c(d)}(q,k) + d_{-q\eta}V_\eta^{c(d)}(q,k) + \tilde{d}_{-q\eta}\widetilde{V}_\eta^{c(d)}(q,k)$$

$$= N \sum_{c'=b,f} \sum_{\tilde{\alpha}_{c'}} \sum_{k'} \epsilon_{c'} F_{k'-\frac{q}{2}}^{c'} \left[ i\omega V_0^{c'(b)}(q,k') + d_{-q\eta}V_\eta^{c'(b)}(q,k') + \tilde{d}_{-q\eta}\widetilde{V}_\eta^{c'(b)}(q,k') \right] \quad (A8)$$

$$\times F_{k'+\frac{q}{2}}^{c'} e^{-i\frac{q_{\eta'}}{2}} m_{\alpha\eta'\mathbf{k}'} W_{\alpha\eta'}^c(q,k)$$

(compare with the last Eq. (5.21)). Using here (A7) and shifting the summation index, we can rewrite this expression as

$$\frac{1}{2}N \sum_{c'=b,f} \sum_{\tilde{\alpha}_{c'}} \sum_{k'} \epsilon_{c'} e^{-i\frac{q_\eta}{2}} \left( m_{\alpha\eta\mathbf{k}'-\mathbf{q}/2}\widetilde{\tau}_\alpha - \widetilde{\tau}_\alpha m_{\alpha\eta\mathbf{k}'+\mathbf{q}/2} \right) F_{k'}^{c'} W_{\alpha\eta}^c(q,k)$$

$$= -\frac{il}{2}N \sum_{c'=b,f} \sum_{\tilde{\alpha}_{c'}} \sum_{k'} \epsilon_{c'} \left( \tilde{d}_{-q\eta}\tilde{g}_\alpha + d_{-q\eta}j_\alpha \right) m_{\alpha\eta\mathbf{k}'} F_{k'}^{c'} W_{\alpha\eta}^c(q,k), \quad (A9)$$

where we have used (A1). Finally, using the self-consistency equations, the last expression may be written as

$$-\frac{il}{2}N \sum_\alpha \left( \tilde{d}_{-q\eta}\tilde{g}_\alpha + d_{-q\eta}j_\alpha \right) M_{\alpha\eta} W_{\alpha\eta}^c(q,k). \quad (A10)$$

Since this is just equal to minus the contribution of the diagram Fig. 4(c) to Eq. (6.22), our proof is completed.